\documentclass[fleqn,12pt]{wlscirep}

\usepackage[utf8]{inputenc}
\usepackage[T1]{fontenc}

\usepackage{verbatim}
 \usepackage{setspace}
\usepackage{soul}
\soulregister\cite7
\soulregister\ref7

\usepackage{etoc}

\usepackage{seqsplit}
\usepackage{gensymb}
\usepackage{dirtytalk}
\usepackage{outlines}
\usepackage{esdiff}
\usepackage{csquotes}
\usepackage{adjustbox}

\usepackage{amsmath}
\usepackage{verbatim}

\title{Diverging network architecture of the \textit{C. elegans} connectome and signaling network}

\author[1]{Sophie Dvali}
\author[2,3]{Caio Seguin}
\author[4,5]{Richard Betzel}
\author[1,6*]{Andrew M. Leifer}
\affil[1]{Princeton University, Department of Physics, Princeton, NJ,  United States of America}
\affil[2]{University of Melbourne and Melbourne Health, Melbourne Neuropsychiatry Centre, Melbourne, Victoria, Australia}
\affil[3]{Indiana University, Department of Psychological and Brain Sciences, Bloomington, IN, USA}
\affil[4]{University of Minnesota, Department of Neuroscience, Minneapolis, MN, USA}
\affil[5]{Masonic Institute for the Developing Brain, Department of Neuroscience, Minneapolis, MN, USA}
\affil[6]{Princeton University, Princeton Neurosciences Institute, Princeton, NJ, United States of America}

\affil[*]{leifer@princeton.edu}

\newpage
\clearpage

\begin{abstract}
The connectome describes the complete set of synaptic contacts through which neurons communicate. While the architecture of the \textit{C. elegans} connectome has been extensively characterized, much less is known about the organization of causal signaling networks arising from functional interactions between neurons. Understanding how effective communication pathways relate to or diverge from the underlying structure is a central question in neuroscience. Here, we analyze the modular architecture of the \textit{C. elegans} signal propagation network, measured via calcium imaging and optogenetics, and compare it to the underlying anatomical wiring measured by electron microscopy.  Compared to the connectome, we find that signaling modules are not aligned with the modular boundaries of the anatomical network, highlighting an instance where function deviates from structure. However, we find that some of the most striking features of the anatomical network are preserved, as exemplified by the pharynx, which is delineated into a separate community in both anatomy and signaling. We analyze the cellular compositions of the signaling architecture and find that its modules are enriched for specific cell types and functions, suggesting that the network modules are neurobiologically relevant. Lastly, we identify a "rich club" of hub neurons in the signaling network. The membership of the signaling rich club differs from the rich club detected in the anatomical network, challenging the view that structural hubs occupy positions of influence in functional (signaling) networks. The only overlap between the two rich clubs is given by neurons AVEL/R, which have some of the highest degrees in the anatomical network, again illustrating the preservation of the most pronounced features of the network. Our results provide new insight into the interplay between brain structure, in the form of a complete synaptic-level connectome, and brain function, in the form of a system-wide causal signal propagation atlas.
\end{abstract}
\begin{document}

\flushbottom
\maketitle
\thispagestyle{empty}

\clearpage

\section*{Introduction}

How the structure of a biological system supports its functional repertoire is a central question in biology. A defining feature of this question is that it seeks to relate two lenses into the same biological system.
In neuroscience, the relevant structural lens is the network map of synaptic connections in the brain, called a  ``connectome'', which is typically measured using electron microscopy and captures the wiring of the nervous system.  The functional lens reports how the activity of one neuron in the brain either relates to or influences the activity of others. These can be measured by functional MRI, electrophysiology, or calcium imaging. In neuroscience, going back to Cajal \cite{cajal_1909}, there is a strong expectation that structure should inform function \cite{felleman_1991, passingham_2002, van_essen_1998}.

Network science provides a powerful toolkit to characterize and distill key properties of networks \cite{bullmore2009, bassett2017network} that, in some cases, can help link structure to function.  For example, in a simple model of disease spread, the characteristic path-length of the static network of human interactions (structure) has a simple scaling relation with the time to global infection (function) \cite{watts_strogatz1998smallworld, curto_2019_nonlineardynamicsreview}.

The success of examples like these has led to a common implicit hypothesis: that since network properties capture fundamental features of a network, they might persist across network descriptions of both the structural and functional lenses of the brain. This hypothesis is explicit in neuroscience investigations into how anatomy relates to fMRI or EEG recordings \cite{vazquez_rodriguez_2019, seguin2023communication}. It also motivates network investigations into connectomes of the simple nematode \textit{C. elegans} \cite{white_structure_1986, varshney_2011,  yan_2017_barabasi, yemini_2021_neuropal, Uzel_2022} and of Drosophila \cite{betzel_dros_2023, lin_2024}.

However, it remains unclear whether features of the network architecture should be preserved in both a network description of a brain's connectome and a network description of its neural function or dynamics. Many common network features came about because they are useful in the context of specific dynamical assumptions, e.g., small worldness is particularly informative for describing systems with simple linear dynamics, such as disease transmission or random walks \cite{curto_2019_nonlineardynamicsreview}. However, the brain has more complex nonlinear dynamics, and there, the applicability of many of these metrics is less obvious \cite{curto_2019_nonlineardynamicsreview}. Similarly, other properties of dynamics can lead to differences in network descriptions of structure and function. For example, if neural signals rapidly propagate deep through the network across multi-hop paths,  such  ``polysynaptic'' signaling would contribute towards differences in features of the structural and functional networks \cite{seguin2023communication, seguin_2023}. While studies in humans are inherently limited in their scale or resolution, there is evidence that network properties can differ, sometimes dramatically, between network descriptions of correlations in BOLD activity and in gross anatomy, including measures of centrality \cite{power2013evidence, hagmann2008mapping} and community structure \cite{power2011functional,suarez_2020,sporns_2016}.

The extent of agreement between network descriptions of a connectome and neural activity may also depend on how well the wires of the connectome themselves capture all the relevant information for governing neural dynamics. While synaptic transmission is clearly important for neural signaling, there may be other factors not visible from the wires that also contribute. For example, wired connections may exist without a corresponding functional connection because key receptors are absent or downregulated. Gene expression or neuromodulators may alter the number,  mix, or physiological properties of receptors or transmitters to transiently activate subsets of wired connections from a menu of all possibilities   \cite{marder2012neuromod, harriswarrick1991neuromod}. Similarly, there is evidence of functional connections that exist in the absence of wiring. Some of these can be attributed to extrasynaptic signaling, where one neuron releases peptides or other transmitters via dense core vesicles that diffuse through the extracellular milieu to activate other neurons and therefore would not be visible in wiring \cite{lim2016RID,randi_signalprop_2023, taylor_2021, ripoll_sanchez_2023, beets_2023, smith_2019peptides, jekely_2013a_peptides, thiel_2024_peptides}. 

With few exceptions \cite{creamer2024, Uzel_2022, yemini_2021_neuropal}, the link between the synapse-level wiring of a brain's connectome and its function has not been explored in detail and at scale.
Therefore, the extent to which network properties of a brain's connectome match those of a network description of its dynamics remains unclear, especially at brain scale and cellular level. One challenge has been a lack of synapse-level maps of anatomical wiring as well as corresponding causal maps of how one neuron's activity influences another's.

Recently, we measured a brain-wide causal ``Signal Propagation Atlas'' of the nematode \textit{C. elegans} \emph{via} direct optogenetic activation and simultaneous whole-brain calcium imaging \cite{randi_signalprop_2023}. Critically, this signaling network is accompanied by a complete connectome of chemical synapses and gap junctions between pairs of neurons \cite{white_structure_1986, cook_whole-animal_2019, witvliet_connectomes_2021}. This gives us the unique opportunity to directly compare network features of the physical wiring of a nervous system with its causal signal flow empirically for the first time at this scale and resolution.

Here, we investigate and compare the properties of both networks: wiring and signaling, focusing on their community structure and rich club. We show that for these network properties, only the most striking features are preserved, and otherwise, there are significant differences between the two networks. For example, both networks have a distinct community corresponding to the pharynx, but most other communities do not correspond across the two networks. Similarly, the most connected member of the rich club of the signaling network is also the most connected member of the connectome, but there is little other overlap in rich club membership.

Collectively, our findings suggest that the effective signaling network, which measures the collective influence of direct, indirect, and extrasynaptic signaling pathways, has different network properties from the underlying connectome.

\section*{Results}

Here, we compare the network properties of the \textit{C. elegans} connectome \cite{white_structure_1986, cook_whole-animal_2019, witvliet_connectomes_2021} --the network of chemical and electrical synapses of the brain-- to a recently-constructed network map of neural signaling \cite{randi_signalprop_2023} of the genetically identical brain.  
While the connectome encodes only direct synaptic connections between neurons, the signaling network captures effective connections that arise from the combined effect of not only direct connections but also all interactions that span multiple hops through the anatomical network (poly-synaptic chains), as well as extrasynaptic signaling that is not constrained to flow along synapses.  We consider binarized directed connectome and signaling networks of the same head neurons ($N = 188$ head neurons from a total of 302 neurons in the entire nervous system), but each having different numbers of edges (corresponding to $M = 1151$ for signaling and $M = 3250$ for the connectome). Binarization of the anatomical network was performed by counting the average number of connections across the four connectomes \cite{white_structure_1986, witvliet_connectomes_2021} (see Methods). By default, a single synaptic contact in any of the four connectomes was sufficient for inclusion as an edge in the network. For the signaling network, pairs of neurons that exhibited a calcium response with a q-value of less than 0.05 were included.  Thus, the networks differ in terms of their density ($d_{s} = 0.033$ and $d_{a} = 0.092$) and their distribution of motifs (Extended Data Fig. \ref{fig:triads}).  The signaling network has an average clustering coefficient of $0.13$, a bit less than half of the anatomical network's clustering coefficient of $0.3$. The \textit{C. elegans} connectome is well known to be a small world network \cite{watts_strogatz1998smallworld, varshney_2011}. Similarly, we find that the signaling network is also small world ($\sigma_{\text{s}} = 4.76$ and $\sigma_{\text{a}} = 2.52$ for the signaling and anatomical head networks, respectively).

\subsection*{Hierarchical community structure differs between signal propagation and anatomy}
\begin{figure}[htbp]
\centering
\includegraphics[width=\linewidth]{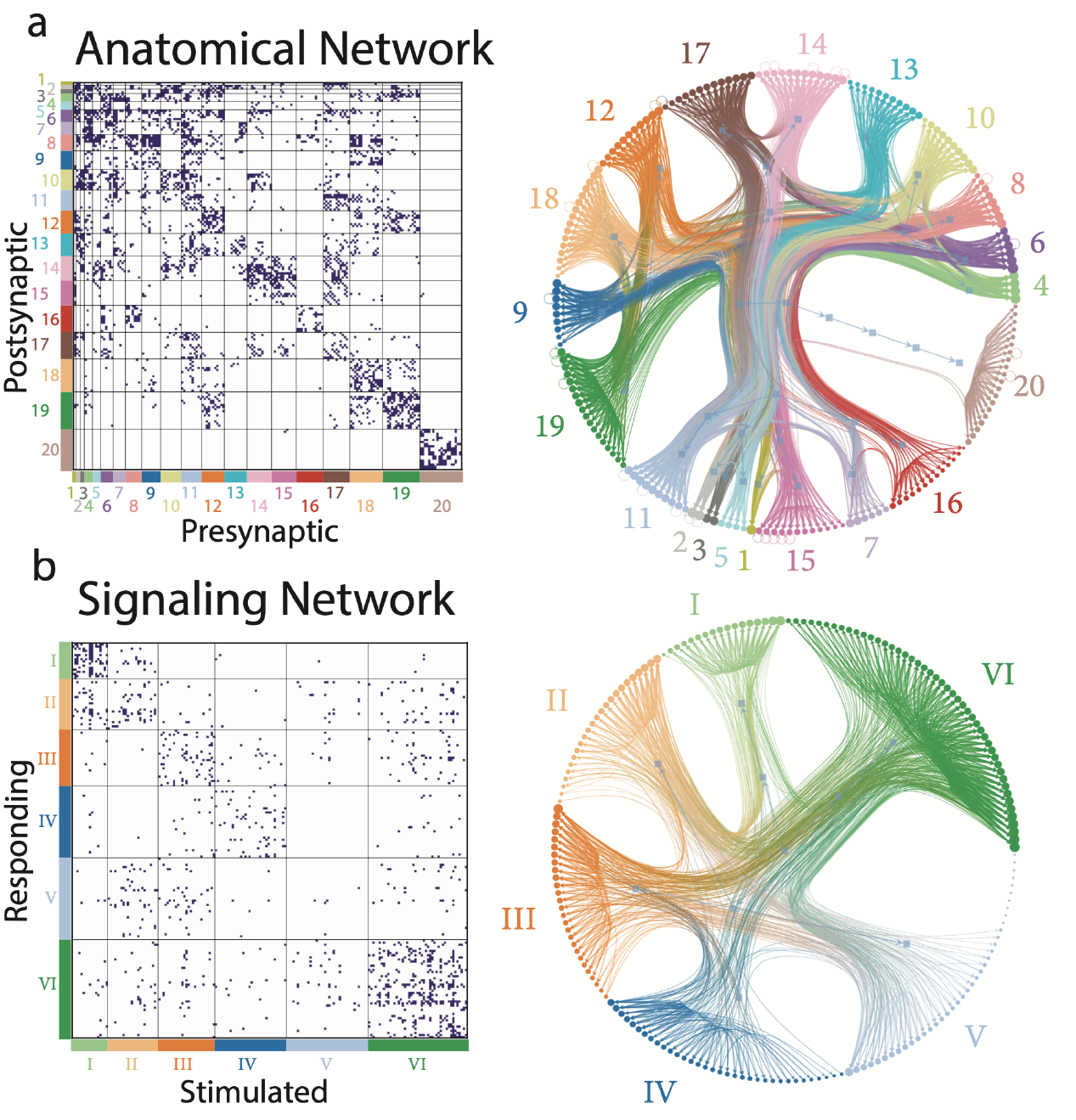}
\caption{\textbf{Hierarchical modular structure of the \textit{C. elegans} signal propagation and anatomical networks.} \textbf{a)} Left: matrix showing anatomical connectivity. Each pixel indicates the presence or absence of synaptic contacts from one individual neuron in the brain to another, from \cite{white_structure_1986, cook_whole-animal_2019, witvliet_connectomes_2021}. Neurons are sorted by community assignment of the first hierarchical level, the smallest community first, and then subsequent communities increasing in size. Right panel: circular dendrogram showing community assignments at different hierarchical levels, colored corresponding to the first level. \textbf{b)} Left panel: signal propagation network matrix. Each pixel indicates whether a downstream neuron exhibits robust calcium activity in response to an upstream neuron's optogenetic stimulation \cite{randi_signalprop_2023}. Organized similarly to (a). Right panel: same as in (a) for the signal propagation network. }  \label{fig:heirarch}
\end{figure}

We chose to compare the complex architecture of wiring and signaling networks by means of their hierarchical modular structure. Neural networks are hierarchically divided into sub-networks, referred to as modules, communities, or clusters \cite{betzel_2017, jarrell2012connectome, betzel2020organizing, betzel2019stability, kiyooka2023single, lee2016anatomy}. This modular composition is thought to shape neural communication by providing a balance between segregation into specialized functional units and integration of information across distant neural elements \cite{sporns_2016} and therefore informs how the nervous system performs computations\cite{suarez_2021, tanner_2023, zhang_2024}. In fMRI studies, abnormalities in modular structure serve as biomarkers of neuropsychiatric conditions \cite{lynch2024frontostriatal}. 
Modular structure has been used to interpret both anatomical \cite{sporns_2013, sporns_2005, sporns_2016} and functional  \cite{power2011functional, bassett_2011reconfiguration} representations of neural networks. Yet, it remains unclear whether modular structure should be conserved in these different views, and in some cases, evidence suggests they are not \cite{suarez_2020, betzel2013multi}.

We explicitly tested whether the anatomical and signaling descriptions of the \textit{C. elegans} brain shared a similar modular structure. We used a hierarchical variant of the stochastic block model (SBM) to independently subdivide the wiring and signaling networks into nested communities \cite{peixoto2014graph, peixoto2014hierarchical, peixoto2019bayesian, karrer2011stochastic, holland1983stochastic, onuchin_2023SBM}. The anatomical and signaling networks differ in the number of communities and in their hierarchical organization. The anatomical network had more communities, it was optimally partitioned into 20 communities in the first hierarchical level, with the smallest containing only one neuron and the largest containing 20 (Fig.~\ref{fig:heirarch} b, anatomical clusters are denoted by Arabic numerals). In contrast, the signal propagation network had fewer and was optimally partitioned into six communities in the first hierarchical level, the smallest of which contained 17 neurons and the largest of which contained 47 neurons (Fig.~\ref{fig:heirarch} a, signal propagation clusters are denoted by Roman numerals).  The larger number of communities in the organization of the anatomical network was not trivially explained by its higher density. We performed two comparisons using different versions of the signaling and anatomy networks, in which we enforced the network density to be the same by adjusting the threshold for inclusion of edges of either network, called ``enforced density controls.'' In one case, we compared the existing signaling network to a more conservative anatomical network in which we required neurons to have more synaptic contacts for inclusion. In another case, we compared the existing anatomical network to a more permissive signaling network that included lower signal-to-noise responses. In both sets of comparisons, the anatomical network had more communities (Extended Data Fig. \ref{fig:aconn_thresh}, \ref{fig:fconn_thresh}). This difference was also robust to the use of a more stringent thresholding-- a ``conservative control''  in which both had more restrictive inclusion criteria but different densities (Extended Data Fig. \ref{fig:conservative_thresh}).

\begin{table}[]
\centering
\begin{tabular}{|l|l|}
\hline
Community I   & \begin{tabular}[c]{@{}l@{}}I2L, I2R, I3, I4, IL2R, M1, M2L, M2R, M3L, M3R, M4, MCL, MCR, MI,\\ NSML, NSMR, URYVR\end{tabular}                                                                                                                                                                                                                        \\ \hline
Community II  & \begin{tabular}[c]{@{}l@{}}AVL, BAGL, BAGR, CEPVL, I1L, I1R, I6, IL1DL, IL1DR, IL1VL, IL1VR,\\ IL2DR, IL2VR, M5, OLLL, OLLR, OLQDR, OLQVR, RIPR, RMER, \\ SAADL, URBR, URYDL, URYVL\end{tabular}                                                                                                                                                     \\ \hline
Community III & \begin{tabular}[c]{@{}l@{}}AFDL, AIML, AIMR, AIYL, AIYR, ASJL, AUAL, AUAR, AVBL, RICL, \\ RIH, RIML, RIMR, RMDDR, RMDL, RMFL, RMFR, RMGL, SABVL, \\ SABVR, SIBVR, SMBDL, SMBVL, SMBVR, SMDDL, SMDDR, VB2\end{tabular}                                                                                                                                \\ \hline
Community IV  & \begin{tabular}[c]{@{}l@{}}ADAL, ADAR, ADEL, ADER, AIBR, AINL, AIZL, AIZR, AQR, AS1, \\ AVFL, AVFR, AVG, AVKL, AVKR, DB1, DB2, DD1, FLPL, FLPR, RIBL, \\ RIBR, RICR, RIFR, RIGL, RIGR, RIS, RMHL, SABD, SIBVL, SMBDR, \\ URBL, VA1, VB1\end{tabular}                                                                                                 \\ \hline
Community V   & \begin{tabular}[c]{@{}l@{}}ADFR, AIAL, AIAR, AIBL, ALA, AWCL, CEPVR, DA1, I5, IL1L, IL1R, \\ IL2DL, IL2L, IL2VL, OLQDL, OLQVL, RID, RIFL, RIPL, RIR, RMDDL, \\ RMED, RMEL, RMEV, RMGR, SAADR, SIADL, SIADR, SIAVL, SIAVR, \\ SIBDL, SIBDR, URADL, URADR, URAVL, URAVR, URYDR, VD1, VD2\end{tabular}                                                  \\ \hline
Community VI  & \begin{tabular}[c]{@{}l@{}}ADFL, ADLL, ADLR, AFDR, AINR, ASEL, ASER, ASGL, ASGR, ASHL, \\ ASHR, ASIL, ASIR, ASJR, ASKL, ASKR, AVAL, AVAR, AVBR, AVDL, \\ AVDR, AVEL, AVER, AVHL, AVHR, AVJL, AVJR, AWAL, AWAR, AWBL, \\ AWBR, AWCR, CEPDL, CEPDR, RIAL, RIAR, RIVL, RIVR, RMDR, RMDVL, \\ RMDVR, SAAVL, SAAVR, SMDVL, SMDVR, URXL, URXR\end{tabular} \\ \hline
\end{tabular}
\caption{\textbf{Community assignments of the signaling network (first hierarchical level).}}
\label{tab:signaling_communities}
\end{table}

\begin{table}[]
\centering
\begin{tabular}{|l|l|}
\hline
Community 1  & RIH                                                                                                                                                   \\ \hline
Community 2  & RIBL, RIBR                                                                                                                                            \\ \hline
Community 3  & RIAL, RIAR                                                                                                                                            \\ \hline
Community 4  & AIBL, AIBR, RIML, RIMR                                                                                                                                \\ \hline
Community 5  & URYDL, URYDR, URYVL, URYVR                                                                                                                            \\ \hline
Community 6  & AVEL, AVER, RIGL, RIGR, RIS, RMFL                                                                                                                     \\ \hline
Community 7  & AVKL, AVKR, SAADL, SAADR, SAAVL, SAAVR                                                                                                                \\ \hline
Community 8  & AVAL, AVAR, AVBL, AVBR, AVDL, AVDR, AVL, RID                                                                                                          \\ \hline
Community 9  & ADAL, ADAR, ALA, AQR, AVG, AVJL, AVJR, FLPL, FLPR                                                                                                     \\ \hline
Community 10 & RIVL, RIVR, RMDL, RMDR, RMFR, SIAVR, SMDDL, SMDDR, SMDVL, SMDVR                                                                                       \\ \hline
Community 11 & ADEL, ADER, RICL, RICR, RMGL, RMGR, URBL, URBR, URXL, URXR                                                                                            \\ \hline
Community 12 & ADFL, ADFR, AIZL, AIZR, AUAL, AUAR, AWBL, AWBR, BAGL, BAGR, RIR                                                                                       \\ \hline
Community 13 & \begin{tabular}[c]{@{}l@{}}SIADL, SIADR, SIAVL, SIBDL, SIBDR, SIBVL, SIBVR, SMBDL, SMBDR, SMBVL,\\ SMBVR\end{tabular}                                 \\ \hline
Community 14 & \begin{tabular}[c]{@{}l@{}}IL1L, IL1R, OLLL, OLLR, RMDDL, RMDDR, RMDVL, RMDVR, RMED, RMEL,\\ RMER, RMEV\end{tabular}                                  \\ \hline
Community 15 & \begin{tabular}[c]{@{}l@{}}IL1DL, IL1DR, IL1VL, IL1VR, IL2VL, IL2VR, RIPL, RIPR, URADL, URADR,\\ URAVL, URAVR\end{tabular}                            \\ \hline
Community 16 & AS1, DA1, DB1, DB2, DD1, SABD, SABVL, SABVR, VA1, VB1, VB2, VD1, VD2                                                                                  \\ \hline
Community 17 & \begin{tabular}[c]{@{}l@{}}CEPDL, CEPDR, CEPVL, CEPVR, IL2DL, IL2DR, IL2L, IL2R, OLQDL, OLQDR,\\ OLQVL, OLQVR, RMHL\end{tabular}                      \\ \hline
Community 18 & \begin{tabular}[c]{@{}l@{}}ADLL, ADLR, AIML, AIMR, ASHL, ASHR, ASJL, ASJR, ASKL, ASKR, AVFL, \\ AVFR, AVHL, AVHR, RIFL, RIFR\end{tabular}             \\ \hline
Community 19 & \begin{tabular}[c]{@{}l@{}}AFDL, AFDR, AIAL, AIAR, AINL, AINR, AIYL, AIYR, ASEL, ASER, ASGL, ASGR, \\ ASIL, ASIR, AWAL, AWAR, AWCL, AWCR\end{tabular} \\ \hline
Community 20 & \begin{tabular}[c]{@{}l@{}}I1L, I1R, I2L, I2R, I3, I4, I5, I6, M1, M2L, M2R, M3L, M3R, M4, M5, MCL, MCR, MI, \\ NSML, NSMR\end{tabular}               \\ \hline
\end{tabular}
\caption{\textbf{Community assignments of the anatomical connectome (first hierarchical level).}}
\label{tab:aconn}
\end{table}

Next, we investigated the distribution of bilateral pairs across communities. A majority of \textit{C. elegans} neurons consist of bilaterally symmetric pairs that are often connected to each other via gap junctions and tend to have similar gene expression  \cite{taylor_2021}, correlated activity patterns \cite{Uzel_2022}, and often (but not always \cite{birari_2024bilateral}) similar wiring \cite{white_structure_1986}. As we would expect, bilaterally symmetric neurons are much more likely to be found in the same communities than across different communities, for both the anatomical and signaling networks. In the anatomical network, bilateral neuron pairs were 38 times more likely to be in the same community than in different communities  ($p = 1.5 \times 10^{-210}$) and in the signaling network, they were twice as likely ($p = 4 \times 10^{-45}$). Even when adjusting thresholds for inclusion in the networks, this trend persisted: two bilaterally symmetric neurons were more likely to be clustered in the same versus different communities, and the degree of enrichment was higher for anatomy than for signaling (Extended Data Table \ref{tab:LR_table}). This finding is consistent with the hypothesis that some features should be conserved across both network descriptions.

\subsection*{Anatomical and signal propagation communities have mostly non-overlapping membership}

\begin{figure}[htbp]
\centering
\includegraphics[page=1, width=.75\linewidth]{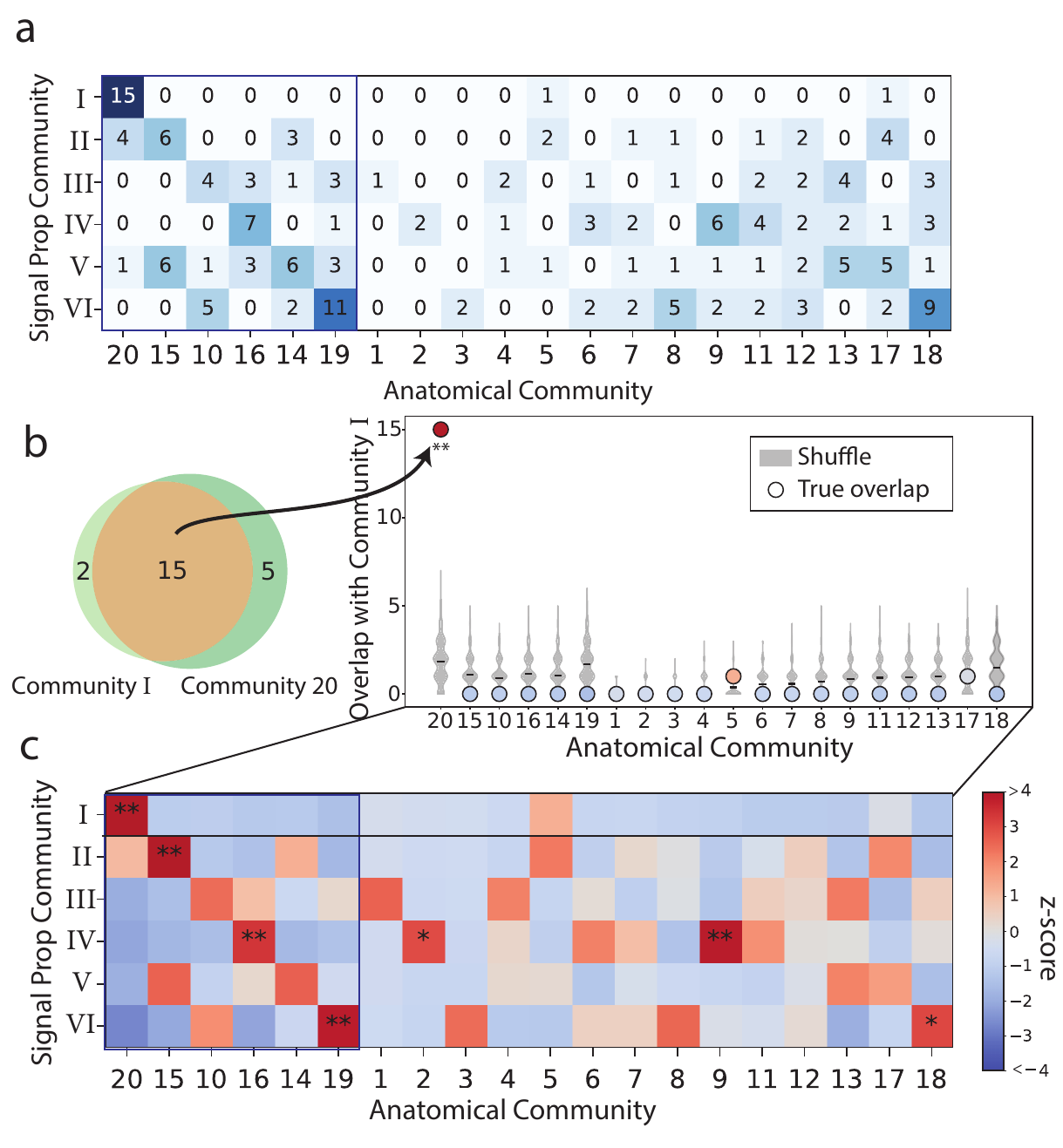}
\caption{ \textbf {Comparison between anatomical and signal propagation communities. a)} Overlap (counts) between membership of anatomical and signal propagation communities is shown. Community best-matching is performed via a similarity matrix and the Hungarian method, as described in the methods.   
\textbf{b)}  Left: Venn diagram illustrating the true overlap (intersection) between the set of neurons in signaling community I and the set of neurons in the anatomical community 20. Right: Community I's number of overlapping neurons for each anatomical module (circles).  Relative enrichment (circle color) is calculated as a z-score of the true overlap compared to a null distribution of shuffled community assignments (grey violin plots), described in methods. Circle color corresponds to color bar in c.  
\textbf{c)} Relative enrichment of overlap is reported as a z-score for each anatomy-signaling community pair  (* $p<0.05$, ** $p<0.01$, p-values from same distribution as z-score, FDR-adjusted). }  \label{fig:aconn_vs_funconn}
\end{figure}

If the community architecture of the signaling and anatomical networks is similar, we would expect signal propagation communities to either correspond to specific partners in the anatomical network, or, since there are fewer signaling communities, we would expect some signaling communities to entirely contain anatomical communities. To test whether signaling and anatomical communities follow this straightforward correspondence, we first calculated the pairwise overlap between each of the signal propagation and anatomical communities (Fig.~\ref{fig:aconn_vs_funconn} a). Signal propagation community \textbf{I} had large overlap with Anatomical Community 20 both of which  correspond to the pharyngeal subnetwork. Other than the pharynx, signal propagation communities did not neatly map onto or wholly contain anatomical communities (adjusted rand score $ARS = 0.105$). 
Most signal propagation communities had neuron membership that were spread across multiple anatomical communities (rows of Fig.~\ref{fig:aconn_vs_funconn} a). Similarly, most anatomical communities had neuron membership that was spread across many signal propagation communities  (columns of Fig.~\ref{fig:aconn_vs_funconn} a). 

We further tested whether the amount of overlap we do observe could be explained by chance. We therefore calculated the extent to which each signal propagation community was enriched for anatomical communities, and compared this to a null hypothesis in which neurons are assigned to communities at random \cite{seguin_2022}, (Fig.~\ref{fig:aconn_vs_funconn} b). Of all possible signal-anatomy community pairings only seven were significantly enriched, considering multiple hypothesis testing.
Overlap was not improved by altering the stochastic block model algorithm to force it to partition each network into the same number of communities  ($ARS_{same6} = 0.002$ and $ARS_{same20} = 0.014$; Extended Data Fig. \ref{fig:same_blocks}).

These findings suggest that most signal propagation communities are not trivially mapped onto nor trivially contain anatomical communities.
The pairing of signaling community \textbf{I} and anatomical community \textbf{20} stands out as an exception. The correspondence between pharyngeal communities in signaling and anatomy persists even in our conservative and enforced density controls (Extended Data Fig. \ref{fig:conservative_thresh}, \ref{fig:aconn_thresh}, \ref{fig:fconn_thresh}).

Why might pharyngeal communities neatly correspond, when others do not? The pharynx has long been hypothesized to be largely functionally autonomous \cite{avery_2013pharynx,albertson_1976pharynx}, partly because it makes only two extrapharyngeal synaptic connections to the rest of the network (Fig.~\ref{fig:heirarch} a) \cite{white_structure_1986, cook_whole-animal_2019, witvliet_connectomes_2021}. Its role in controlling pharyngeal pumping and feeding behaviors has historically been assumed to operate largely independently of the rest of the network \cite{albertson_1976pharynx, avery_2013pharynx}. This stark anatomical isolation may explain why both the structural and signaling networks preserve this community organization. It is interesting that, even with this direct correspondence between the anatomical and signaling communities of the pharyngeal neurons, in the signaling network, the community of pharyngeal neurons is much more highly connected to the other communities than it is in the anatomical network (Fig.~\ref{fig:heirarch} a), highlighting differences between the brain's wiring and signaling. 

Collectively, our results suggest that, while the most striking communities may be preserved, in general, modular structure is not conserved across anatomical and functional descriptions of the brain. In the discussion, we speculate that a combination of polysynaptic signaling, extrasynaptic signaling, and non-functional anatomical wiring may contribute to these differences. The pharynx suggests that in cases where a sub-network of neurons is extremely isolated in the anatomical network, its modularity may be preserved even in the functional description. However, even here, it is notable that the patterns of connectivity between the modules still differ.

\subsection*{Communities are enriched for neuron type and role}

\begin{figure}[htbp]
\centering
\includegraphics[page=1, width=0.7\linewidth]{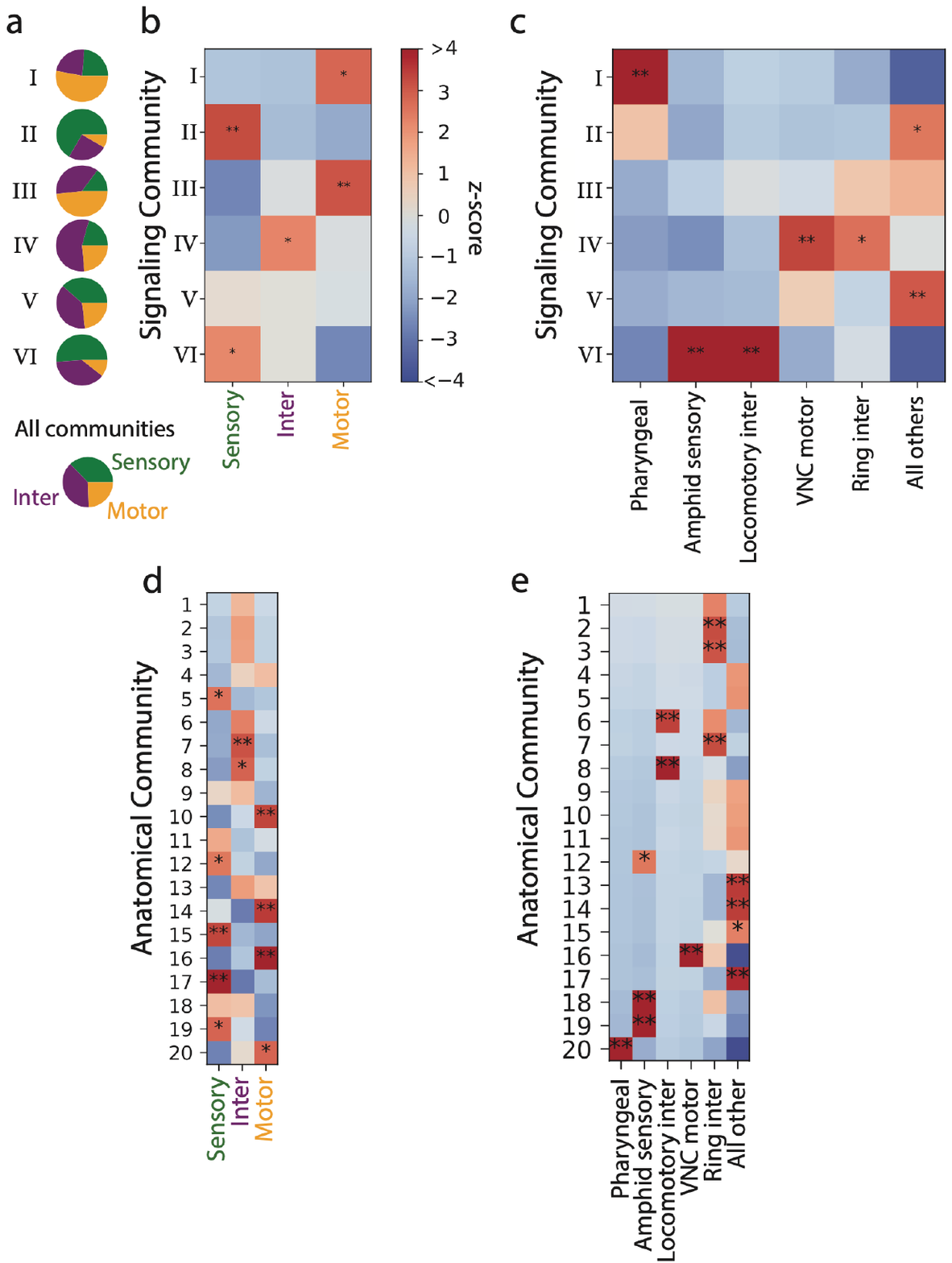}
\caption{ \textbf {Communities are enriched for neuron type and role. a)} Fractional breakdown of signaling community neuron membership into sensory, inter-, and motor neurons cell type (green, purple, and yellow respectively). \textbf{b)} Enrichment (z-score) of signaling communities for sensory, inter-, and motor neurons (* p<0.05, ** p<0.01, p-values FDR-adjusted)  \textbf{c)} Enrichment of signaling communities for neuron role (* p<0.05, ** p<0.01, p-values FDR-adjusted) \textbf{d)} Enrichment (z-score) of anatomical communities for sensory, inter-, and motor neurons (* p<0.05, ** p<0.01, p-values FDR-adjusted), \textbf{e)} Enrichment of anatomical communities for neuron role (* p<0.05, ** p<0.01, p-values FDR-adjusted)}  \label{fig:enrich}
\end{figure}

Clustering of anatomical networks has previously been used to identify communities of neurons that are enriched for cell types or known functional cellular roles, including in \textit{Drosophila} and \textit{C. elegans} \cite{betzel_dros_2023, betzel2024parallel, Emmons_clustering, weinstein2024network, li2024network, kunin2023hierarchical}. Consistent with prior findings,  our modular clustering of the \textit{C. elegans} anatomical network recapitulates anatomical communities that are each significantly enriched for no more than one of the three \textit{C. elegans} neuron cell types classes: Sensory, Inter- and Motor neurons \cite{varshney_2011}(Fig.~\ref{fig:enrich} c; Extended Data Fig.~\ref{fig:enrich_sup_nlist} a; Table \ref{tab:aconn}). Similarly, they are each enriched for no more than one  of five predefined cell roles (pharyngeal, amphid sensory, locomotory inter-, vnc motor and ring interneurons) (Fig.~\ref{fig:enrich} d; 
Extended Data Fig.~\ref{fig:enrich_sup_nlist} b; Table \ref{tab:aconn}). Enrichment is calculated by comparison to a null hypothesis of randomly shuffled assignments as described in methods. We next explored cell-type annotation in the signaling network.  

Signaling network communities, like the anatomical communities, also showed clear enrichment for distinct cell types. Each signaling community was statistically significantly enriched for no more than one of three previously defined cell type classes (Fig.~\ref{fig:enrich} a,b; Extended Data Fig.~\ref{fig:enrich_sup_nlist} a; Table \ref{tab:signaling_communities}).  Only one of the six communities, \textbf{V}, showed no significant enrichment for any cell type.

Similarly, cell roles are well segregated into different signaling communities. As mentioned in the previous section community \textbf{I} contains almost all the pharyngeal neurons. All but one locomotory interneuron are members of community \textbf{VI} (Fig.~\ref{fig:enrich} c; Extended Data Fig.~\ref{fig:enrich_sup_nlist} b; Table \ref{tab:signaling_communities}).
 Along with interneurons involved in locomotion, community \textbf{VI} was also enriched for the amphid sensory neurons. 
 The amphid sensory neurons are the primary olfactosensory neurons of \textit{C. elegans} and play a  role in chemotaxis and mechanosensation \cite{ bargmann_1991_chemotaxis,  bargman_1997_chemotaxis, driscoll_1997_mechanotransduction}. The locomotor interneurons AVA, AVD, and AVE from this community, among others, translate this sensory signal into behavior to control and initiate backward locomotion  \cite{AVE_kumar_2023, AVE_wang_2020, AVE_kawano_2011, gordus_2015, gray_2005}. Similarly, neuron AVB from this community is a key part of the forward locomotory circuit \cite{chalfie_1985_AVB}. The community structure may therefore have defined a cluster in such a way as to capture a sensorimotor pathway.

 Community \textbf{IV}, the only community that was enriched for interneurons, was not only specifically enriched for ring interneurons but also for VNC motor neurons. The neurons that make up the ring interneurons have been less well characterized than the locomotor interneurons but are thought to integrate sensory signals and information about the inner state of the organism to drive behavioral decisions likely through VNC motor neurons which innervate dorsal and ventral muscles to induce locomotion \cite{tsalik_2003_RIB,white_structure_1986, celegansbook}.

That signaling communities are enriched for cell type, and that cell roles are segregated into different communities, sometimes in sensible ways, suggests that our signaling measurements reflect known functional roles in the brain and also provides evidence that the stochastic block model analysis extracts meaningfully defined communities.

\subsection*{Anatomy is net feed-forward along the sensorimotor path while signaling is ambiguous}

It is commonly expected that there should be overall feed-forward signaling from sensory neurons to interneurons to motor neurons \cite{varshney_2011}. Anatomically, synaptic edges are enriched in a feed-forward manner from sensory neurons to interneurons to motor neurons (Extended Data Fig.~\ref{fig:SIM2} b,c). By contrast, the signaling network is not appreciably feed-forward along this path compared to a shuffle control (Extended Data Fig.~\ref{fig:SIM2} b,c). This is also visible in the connectivity between and among communities. We do not see any obvious evidence of feed-forwardness when looking at how communities that are enriched for sensory, inter, and motor neurons signal to each other (Extended Data Fig.~\ref{fig:SIM}).

\subsection*{Linking neurobiological annotations to community motifs}

\begin{figure}[htp]
\centering
\includegraphics[page=1, width=\linewidth]{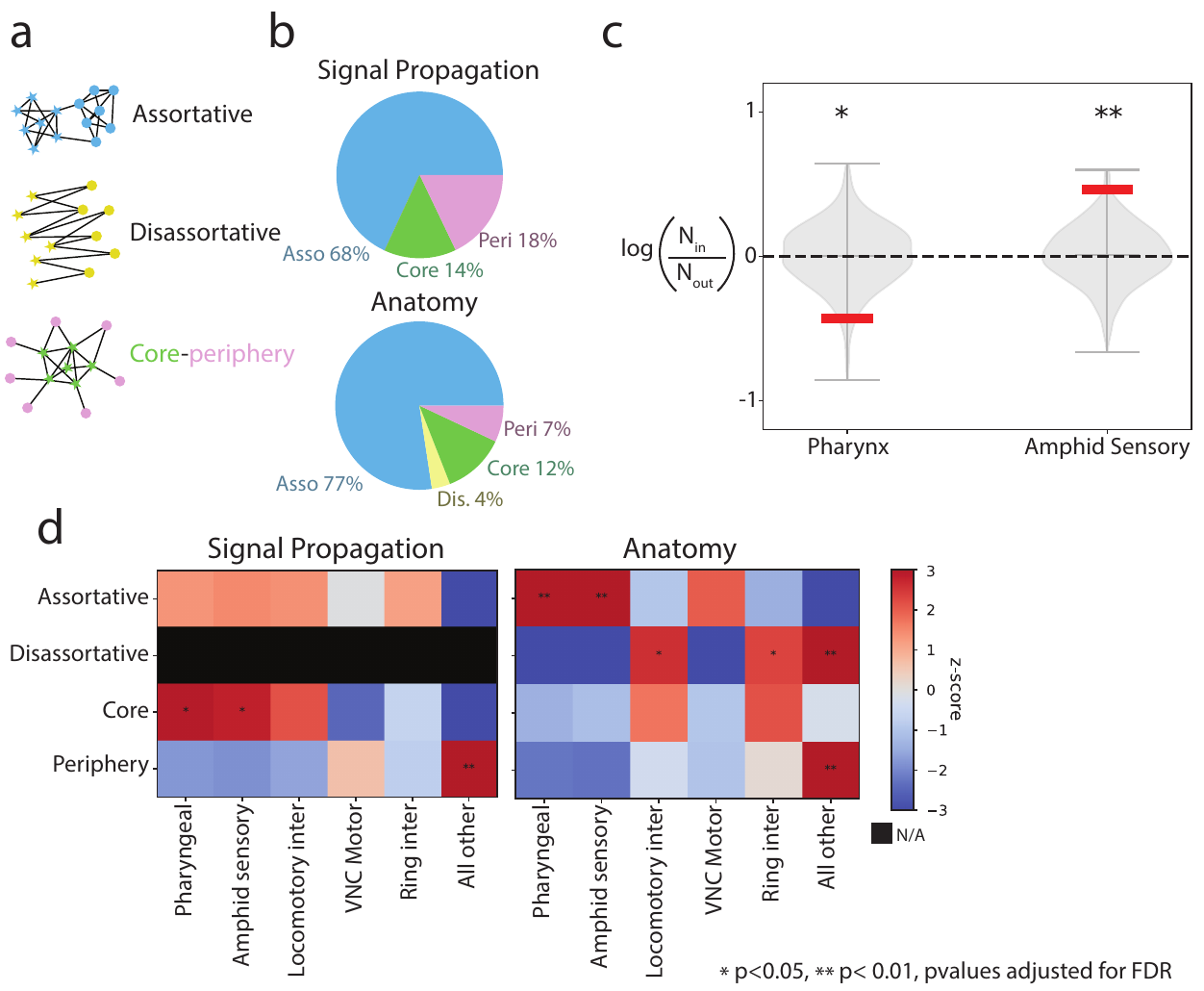}
\caption{ \textbf {The Pharynx and Sensory neurons make up the core of "core-periphery" interactions. a)} Example schematic of Assortative, Disasortative, and Core-periphery community interactions \textbf{b)} Pie charts of the average fraction of participation of each node in the different community interactions for the signal propagation network (Top Panel) and the anatomical network (Bottom Panel). \textbf{c)} Log of the ratio of incoming and outgoing connection to the pharynx and the amphid sensory neurons (red bar) compared to null distribution from networks where the neuron role assignments have been shuffled (grey violin plot) (* p<0.05, ** p<0.01). \textbf{d)} Enrichment (z-score) of neurons of different roles for participation in assortative, disassortative, and core/periphery interactions in the signaling network (Left Panel) and anatomical network (Right Panel). Signaling network has no disassortative interactions. (* p<0.05, ** p<0.01, p-values FDR-adjusted)}  \label{fig:ADCP}
\end{figure}

To gain insight into large-scale network organization, we identified stereotyped inter-community relations, called ``community motifs'' that assign a label to how each community interacts with each other community (Fig.~\ref{fig:ADCP} a) \cite{betzel_human_nonhuman_2018}. Unlike some common methods for community detection designed to detect ``assortative'' community structure (for example, modularity maximization \cite{newman2004finding} or infomap \cite{rosvall2008maps}), the stochastic block model can detect generic types of community structure. 
Specifically, given two communities $r$ and $s$, we considered all community pairs and their within- ($w_{rr}$ and $w_{ss}$, given by the number of edges between nodes in the community divided by the number of possible edges between nodes in the community) and between-community densities ($w_{rs}$ given by the number of edges between nodes in community $r$ and community $s$ divided by the number of possible edges between the two communities), classifying the pairwise interactions as one of the three different community motifs (Figure \ref{fig:ADCP} a). Two communities are classified as assortative if $w_{rr}>w_{rs}$ and $w_{ss}>w_{rs}$. An interaction between communities is disassortative if $w_{rr}<w_{rs}$ and $w_{ss}<w_{rs}$. Lastly, two communities are classified as core-periphery in the case that $w_{rr}>w_{rs}>w_{ss}$.

Different community interactions provide hypotheses about the roles that communities may play when interacting. For example, the presence of the internally dense but externally sparse assortative communities, has been thought to be evidence of separate computational units. We observe, as expected, that the majority of community interactions in both signal propagation and anatomical networks are assortative ($70\%$ for both networks) \cite{betzel_human_nonhuman_2018, betzel_dros_2023, pavlovic2014stochastic}. In both networks, neurons most frequently participated in assortative interactions (Fig. \ref{fig:ADCP} b).

We were particularly interested in investigating core-periphery interactions because they can be evidence of broadcasting and integrating information.  Core-periphery interactions are comprised of one densely connected core of ``driver'' nodes with substantial connections to a periphery that, in turn, is sparsely connected internally. Here in particular we might expect mismatch between signaling and anatomy because, in contrast to anatomy, signaling reports result of many hops of signals through the network, which result in broadcasting or integrating that wouldn't be easily identified from the direct connections of the connectome.
We observed that in the anatomical network, $26\%$ and in the signaling network, $30\%$ of community interactions are core-periphery and we wondered which types of neurons may be involved in this broadcasting and integrating motif.

Strikingly pharyngeal and amphid sensory neurons were significantly enriched for participation in the ``core'' component of the core-periphery interactions in the signal propagation network (Fig.~\ref{fig:ADCP} d). However, in the anatomical network, these same neurons were impoverished for their coreness and instead were significantly enriched for participation in the assortative community motif.  We assessed the enrichment of neurons for participation in these different interaction types, by calculating a z-score comparing the actual interactions participated in to those obtained from randomly shuffled community interactions.  These results show that the divergence between anatomical and signaling modules goes beyond a mere reassignment of nodes into communities, but rather indicates the presence of a differing architecture in the signaling network which may support modes of neural computation and communication that are not evident from examining only the anatomical architecture.

As participation in the core of a core-periphery interaction could potentially indicate a role for broadcasting or integrating signals, we were curious if the pharyngeal and amphid neurons fit one of these two roles. By inspecting the number of ingoing and outgoing edges in the signaling network, we find that pharyngeal neurons show more broadcasting-like behavior perhaps suggesting a role of sending pumping or food-related signals broadly to the rest of the nervous system (Fig.~\ref{fig:ADCP} c). Counterintuitively amphid sensory neurons have more incoming than outgoing edges in the signaling network, possibly suggesting that more research is needed to explore whether they could be serving as integrators (Fig.~\ref{fig:ADCP} c). The ratios of incoming to outgoing edges could not trivially be explained by the number of times in which amphid or pharyngeal neurons were observed or stimulated (Extended Data Fig.~\ref{fig:intbroad} a,b,c,d).

This change in participation in interaction styles of the same group of neurons between the two networks was also observed in other cell roles, including locomotory interneurons which are enriched for participation in disassortative interactions in anatomy, a motif that signifies a capacity for transmitting signals across the boundaries of communities and which is entirely absent from the signaling network.  This switch in interaction styles for the different neuron types between the anatomical network and the signaling network suggests that the functional role of neurons (broadcaster, integrator, etc) may be hard to predict from network analysis of underlying anatomy.

\subsection*{Anatomy and signaling exhibit rich clubs of mostly non-overlapping neurons}

Like  other complex networks \cite{colizza2006detecting,van2011rich}, the \textit{C. elegans} connectome exhibits ``rich-club'' connectivity \cite{towlson_2013}, in which highly connected hub nodes are also densely connected to each other. Hub neurons have been shown to be very relevant for neural dynamics \cite{curto_2019_nonlineardynamicsreview} and therefore are a feature where we might expect an overlap between structure and function. We wondered whether we should also expect to see rich-club features in the signaling network, and which neurons would be involved.

\begin{figure}[htbp]
\centering
\includegraphics[page=1, width=.75\linewidth]{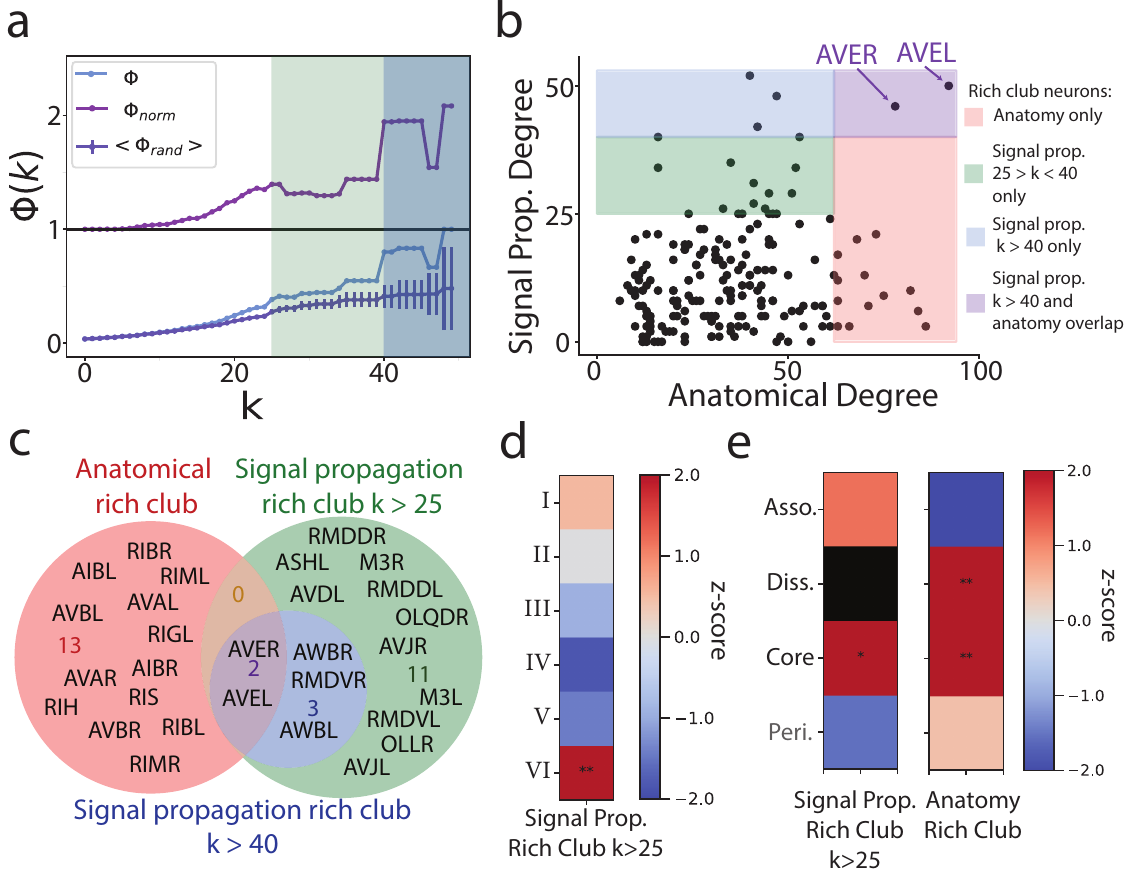}
\caption{ \textbf {Rich club of the signaling network. a)} Normalised rich club coefficient of the signaling network (purple), rich club coefficients of the empirical signaling network (light blue), and 100 randomized null networks (dark blue; errors bars indicate standard deviation). Rich club regimes are marked by green ($k>25$) and blue ($k>40$) shadings.  \textbf{b)} Signaling degree vs. the anatomical degree of each neuron (Pearson correlation $r = 0.22$).  \textbf{c)} Venn diagram of the neurons in the anatomical rich club (red), signaling rich club of $k>25$ (green) and signaling rich club of $k>40$. \textbf{d)} Enrichment (z-score) of the signaling communities for signaling rich club neurons (* p<0.05, ** p<0.01, p-values
FDR-adjusted). \textbf{e)} Enrichment (z-score) of neurons belonging to the signaling rich club (left) and the anatomical rich club (right) for assortative, disassortative, and core/periphery interactions in their respective networks.  }\label{fig:Rich}
\end{figure}

We found statistical evidence supporting the hypothesis that the signaling network also displays the defining feature of rich-club behavior: neurons with a degree of $k$ or higher, highly connected to other neurons of degree $k$ or higher (Fig. \ref{fig:Rich} a). The signaling network exhibited a ``dual rich club'' with two clear tiers: the first tier contained 16 neurons of degree 25 or higher, whereas the second and more exclusive tier contained five neurons of degree 40 or higher (Fig. \ref{fig:Rich} a,b).


There were two neurons present in both rich clubs: command interneurons AVEL and AVER. This bilateral pair of neurons is very highly connected in both networks, which is consistent with its prominent role in the locomotory circuit driving backwards locomotion \cite{AVE_kumar_2023, AVE_kawano_2011, AVE_wang_2020, gray_2005}. This again illustrates that the most extreme features of the network are preserved across anatomy and signaling.
Otherwise, the sets of neurons that comprised the rich clubs in the signaling network surprisingly were nearly disjoint when compared to the rich club in the anatomical network (Fig. \ref{fig:Rich} b,c). Neurons that exhibited the highest degree in the anatomical network tended to have lower degrees in the signaling network (Pearson correlation $r = 0.22$) (Fig. \ref{fig:Rich} b red shading) while neurons with the greatest degree in the signaling network were less well-connected in the anatomical network (Fig.~ \ref{fig:Rich} b blue and green shading). We find that this result persists in both the conservative control and the enforced density controls (Extended Data Fig. \ref{fig:conservative_thresh} d, \ref{fig:aconn_thresh}a, \& \ref{fig:fconn_thresh} a).

We next asked: how are the signaling network's rich-club neurons distributed across its communities? Do they span multiple communities, linking them to one another, or do they predominantly congregate in one community? Here we consider both tiers of the rich club together ($k>25$). Community \textbf{VI} was highly enriched for hub neurons, containing 10 out of the 16 hub neurons, including AVEL and AVER. Rich club neurons also make up a large fraction of this community, approximately 21\%. Given that rich club neurons are, by definition, interconnected, it is then consistent that community \textbf{VI} is also the most highly self-interconnected community (Extended Data Fig. \ref{fig:SIM}). 

Of the remaining rich club neurons, two were in pharyngeal community \textbf{I}  (M3L, M3R), and several were in communities \textbf{II} (OLLR, OLQDR), \textbf{III} (RMDDR) and \textbf{V} (RMDDL). Surprisingly, members of the rich club were fairly equally spread among sensory, inter-, and motor neurons (Extended Data Fig. \ref{fig:Rich_sup} b). This is in contrast to the anatomical network where the rich club is primarily composed of interneurons (Extended Data Fig. \ref{fig:Rich_sup} c). These observations suggest that some neurons historically classified as sensory or motor based on their anatomical wiring may also serve a functional role similar to what has conventionally been ascribed to interneurons.

Due to their nature as highly interconnected and high-degree neurons, we expected that the rich club neurons might function as a core of ``core-periphery" interactions. As predicted, in both the anatomical and signaling networks, each network's respective rich club members were enriched for coreness (Fig. \ref{fig:Rich} e). The interpretation of the core as integrators/broadcasters of information is well-aligned with the known functions of neurons in the anatomical rich club, like AVA, RIM, and AVB, that coordinate motor responses to sensory signals. It is less obvious how well this description fits those neurons in the signaling rich club, such as AVJ, which have not previously been described as sensory integrators. Our findings suggest that it may be worth further investigation to assess whether these neurons have an additional integration role.

Rich club analysis relies on measures of in- and out-degree. A potential confounding factor is that signal propagation degree is correlated with the number of empirical observations made of each neuron in our experimental dataset (pearsons correlation $r =0.71$, Extended Data Fig.~ \ref{fig:intbroad} e). We were reassured, however, that rich club members had large degrees well in excess of what we could explain from the correlation (Extended Data Fig.~\ref{fig:Rich} e,f), and not all highly observed neurons are members. Therefore we conclude that membership in the rich club is not a trivial result of being measured often.

\section*{Discussion}

Here, we take advantage of a recently compiled signal propagation network in \emph{C. elegans} to investigate how the causal network architecture of the worm compares to its underlying wiring, and ultimately how this maps onto cell type and known neurobiological function. In particular, we test whether features commonly used to characterize networks are preserved across these two network descriptions of the same brain. This is important because, until now, only one network description of a brain has generally been available at a time, and therefore understanding how different network descriptions relate is crucial for extracting insights about neural signaling from a wiring diagram, or vice versa.

One's perspective likely influences the degree to which one would expect network features of the anatomical network to be similar to signaling networks. Since they both reflect the same underlying nervous system, it is tempting to assume that key features will be preserved. And indeed, the most obvious features are: the pharynx forms its own community in both anatomy and signaling. In addition, the richest members of the rich club in anatomy are also the richest members of the rich club in signaling. 

However, as we consider the dynamical systems nature of the nervous system, it becomes clear that there are fundamental differences between a network's description of anatomy and a network description of signaling. For example, anatomy describes direct wired interactions, while signaling captures how neural activity travels multi-hop through the network, including feedback, and recurrence.  Even superficially basic assumptions about how to interpret nervous system wiring in a network context are unclear. For example, if a neuron has many anatomical outgoing connections, does this suggest a strong signaling impact because it talks to many post-synaptic patterns, or a weak signaling impact because its synaptic output is diluted across many paths? Network models that consider dilution perform well in humans \cite{seguin_2023} and are one possible avenue for future exploration. Given the challenges, it may not be surprising that network descriptions based on specific assumptions of the underlying dynamics do not necessarily generalize to all brains or even to brains at all \cite{curto_2019_nonlineardynamicsreview}. 

Similarly, as one considers additional biophysical details, it becomes clear that wiring is not the only factor that influences dynamics and that contributions from neuromodulation, extrasynaptic signaling, and gene expression all have important roles to play and are poorly captured by the network description of anatomy alone.

Together, our findings suggest that in real-world brains, it is perhaps only the most pronounced network features,  such as the delineation of the nearly isolated pharyngeal sub-network, that we should expect to be preserved across both signaling and anatomical networks. 

In both signaling and anatomy, we find that bilaterally symmetric neurons are much more likely to be found in the same communities than across different communities, suggesting a strong role of bilateral symmetry in the brain. How do we interpret the observation that this bilateral symmetry is extremely pronounced in the anatomical wiring  (38 times more likely to be in the same community) but less so in signaling (twice as likely)? Here again, fundamental differences in what the networks describe could play a role. Community membership in the anatomical network is influenced entirely by monosynaptic wiring patterns. Therefore, laterality in the wiring patterns one hop away has no bearing on a neuron's community membership in the anatomical network. By contrast, the signaling network reflects poly-synaptic or multi-hop signaling. Therefore, we would expect laterality in wiring more than one hop away to influence a neuron’s signaling community membership. In this way, we expect the signaling network to be generally more susceptible than the anatomical network to laterality in wiring that occurs anywhere in the network. In addition to this increased sensitivity to laterality in wiring, we also expect the signaling network to reflect laterality in gene expression and other non-wired aspects of the nervous system. Therefore, it is perhaps not surprising that we find that the crisp bilaterality of the connectome is most strongly reflected in the anatomical network and only more weakly reflected in the signaling network.

Beyond the striking examples discussed above, nearly all other aspects of the network were found to be different, including the mapping, delineation, and hierarchy of network modules and the composition of each network's respective rich clubs.

Despite the challenges of relating the two networks, we note that the network architecture of the signaling network on its own provides intriguing hypotheses about the possible role of neurons and subnetworks in the brain. 
For example, our analysis finds that the pharynx plays a broadcasting role as the core in ``core-periphery'' interactions between communities, which suggests that it may send out signals related to pharyngeal pumping and food to the rest of the network. This could help explain how other portions of the brain get access to food intake signals to regulate behaviors such as dwelling \cite{rhoades_2019_NSMseratonin}. We find that in the signaling network, the pharynx is less isolated than one would expect from the anatomical network, likely due to polysynaptic signaling and also possibly through extrasynaptic signaling of key pharyngeal neurons, such as M3L \cite{randi_signalprop_2023}.

One surprising finding is that amphid sensory neurons were found to have more incoming than outgoing edges in the signaling network, suggesting that they may play a potential role as integrators, even though they are primarily known for their role in receiving sensory inputs. Another surprising finding is the diversity of neuron types that are found in the signaling rich club, which span sensory, motor, and interneurons. Together, these findings hint at the possibility that the signaling network in \textit{C. elegans} is much more functionally intertwined than previously thought--  neurons are listening to more diverse sets of neural inputs and may take on more diverse roles than previously appreciated.

A challenge of any network analysis, is that it is only as good as the experimental measurements that describe the underlying network. In general, a caveat of measuring signal-evoked responses is that we are limited by the signal-to-noise ratio of our calcium indicator, and we miss weak, noisy functional connections or those of which we have few measurements. For example, in the calcium imaging experiments used to measure the signaling network, some neurons are observed or stimulated more often than others, and this is quantified \cite{randi_signalprop_2023}. Uneven sampling introduces biases, which neurons have edges between them. The signal propagation atlas attempts to account for and correct for this bias by only declaring neurons to be functionally connected after employing a statistical framework that explicitly considers both the magnitude of a neuron's response as well as the number of observations \cite{randi_signalprop_2023}.   Nonetheless, a correlation still persists between the number of times a neuron was observed or stimulated and its in or out degree, respectively, in the signaling network (Extended Data Fig.~\ref{fig:intbroad} a, b, e). Here we present evidence to show that this lingering correlation does not solely explain the richness of the members of the anatomical rich club, for example,  (Extended Data Fig.~\ref{fig:intbroad} e, f) nor can it trivially explain the integrating or broadcasting nature (defined based on the ratio of out to in degree) of the amphid sensory or pharyngeal neurons, respectively (Extended Data Fig. \ref{fig:intbroad} a,b,c,d). In addition, the stochastic block model is implemented in a degree-corrected manner to partition both networks into communities. We also explore several controls in which we alter the thresholds for edge inclusion. Future experimental work is needed to more evenly measure signaling across the network to avoid this confound from the start. 

Inhibition is likely systematically undercounted in the underlying measurements of the signaling network due to technical limitations of measuring calcium activity. Because of this, were inhibition to be organized in a systematic way in the network, we would likely not capture that in our analysis. Finally, both of the networks considered here are binarized approximations. We chose a binarized approach for simplicity, and showed that key results are not sensitive to modest changes in binarization thresholds. However, future work could explore whether a weighted network approach yields more similarities.

A future direction is to try to understand how features of the signaling network may arise from the spatial organization of neurons and synaptic contacts. For example, we expect that extrasynaptic neuropeptide signaling can only occur between pairs of neurons that are sufficiently close so that neuropeptides released from one neuron are sufficiently likely to reach neuropeptide receptors on the other. One challenge is that the locations of peptides and receptors on neurons are not known. More generally, it has been challenging to identify relevant spatial metrics (e.g., distances) between neurons because neurons are extended bodies that often are arranged into dense bundles, and have long processes with extended contact areas across many neighbors \cite{white_structure_1986, witvliet_connectomes_2021}. New progress may make this possible in the near future \cite{koonce_2024_contactome}.

Here we have directly compared properties of signaling and anatomical descriptions of a brain and find that while some properties are conserved in both the anatomical and signaling networks of the nematode \textit{C. elegans}, many other properties are not.  This difference will be important for interpreting network architecture results of newly acquired connectomes \cite{dorkenwald_2024flywire, lin_2024} and of forthcoming connectomes \cite{abbott_2020_mousemind}.  We believe efforts to better understand how these different lenses into the brain relate provide an exciting new frontier into network science \cite{barabasi2023neuroscience}.

\bibliographystyle{naturemag-doi}
\bibliography{NetworkAnalysis_new.bib}

\section*{Acknowledgments} 
Research reported in this work was supported by the National Institutes of Health  National Institute of Neurological Disorders and Stroke under New Innovator award number DP2-NS116768 to A.M.L; the Simons Foundation under award  SCGB \#543003 to A.M.L.; by the National Science Foundation, through the Center for the Physics of Biological Function (PHY-1734030); by the National Science Foundation under award number 2023985 to R.F.B.; by the National Institute of Aging (5R01AG075044-03) to R.F.B; by the National Institute on Drug Abuse (1RF1NS125026-01A1) to R.F.B; by the Boehringer Ingelheim Fonds to S.D.;

\section*{Author contributions}
Conceptualization: S.D., C.S., R.B. and A.M.L.; Formal analysis: S.D.; Funding acquisition: A.M.L.; Investigation: S.D.; Methodology: S.D., C.S. and R.B.; Project administration: A.M.L.; Supervision: A.M.L.; Visualization: S.D.; Writing – original draft: S.D.; Writing - review \& editing: S.D., C.S., R.B. and A.M.L.;

\section*{Competing Interests}
There are no competing interests.

\section*{Additional Information}
To whom correspondence should be addressed: leifer@princeton.edu

\section*{Code availability}
Code available at \url{https://github.com/SophieDvali/FunconnNetworkAnalysis}

\clearpage
\newpage

\section*{Methods}

\paragraph*{The Networks}
The \textit{C. elegans} connectome we use consists of electrical and chemical synapses between pairs of neurons from three adults and one L4 EM datasets \cite{white_structure_1986, witvliet_connectomes_2021}. 
We binarize the connectome in the following way: Two neurons in the anatomical network are considered connected if there is at least one connection between them in any of the four connectome datasets. The signal propagation network was experimentally measured via direct optogenetic activation and simultaneous whole-brain calcium imaging \cite{randi_signalprop_2023}. The signal propagation dataset can be found at \url{https://dandiarchive.org/dandiset/001075/0.240920.1434}\cite{dandi}. Neurons are considered to be statistically significantly functionally connected if they have a q-value of less than 0.05. Both networks contain the 188 head neurons of the 302 total \textit{C. elegans} neurons. 

\paragraph*{Stochastic Block Modeling}
Hierarchical community assignments were determined by finding the best partition using the graph-tool python module agglomerative multilevel Markov chain Monte Carlo (MCMC) algorithm in a degree-corrected manner \cite{peixoto2014graph}\cite{peixoto2014hierarchical}\cite{peixoto2019bayesian} (\url{https://graph-tool.skewed.de/static/doc/demos/inference/inference.html}).  Here, the inferred "best" partition of the communities comes from sampling the equilibrated Markov chain many times, aligning communities, and for each neuron taking the mode of its assignment. Neither networks were weighted. 

\paragraph{Bilaterally symmetric neurons} The probability of two bilaterally symmetric neurons being in the same community and not in the same community were calculated by dividing the number of pairs in the same community and not in the same community respectively by the total number of bilateral pairs. The p-value was determined by comparing the probabilities to a null distribution of probabilities derived from shuffled community assignments via a one-sided t-test. 

\paragraph{Enrichment Analysis}
Community enrichment was determined by calculating the overlap of the set of neurons with a given cell annotation and the set of neurons in a given community (or in the case of comparing the communities of both networks the overlap of the set of neurons in each of the network communities). We then also calculate a null distribution of overlaps for 1000 shuffled community assignments. We then compare the original overlap to the null distribution to determine a z-score and p-value (one-sided t-test) for enrichment. P-values are adjusted for multiple hypotheses using FDR.

\paragraph{Module-matching}
For each pair of communities (anatomical and signal propagation), we calculate the Jaccard score to obtain a similarity matrix. The similarity matrix is then reorganized via the Hungarian method to determine the anatomical communities that best correspond to each of the signal propagation communities.

\paragraph*{Community Motifs Analysis}
To determine if the detected communities in the signaling network exhibited these types of non-assortative interactions (as well as the more common assortative type), we followed a recently-proposed method \cite{betzel_human_nonhuman_2018}.  Specifically, for communities $r$ and $s$, we consider the within-community densities, $\omega_{rr}$ and $\omega_{ss}$, as well as the between-community density, $\omega_{rs}$. Given these values, we unambiguously classified the interaction between $r$ and $s$ as assortative if $\omega_{rr} > \omega_{rs}$ and $\omega_{ss} > \omega_{rs}$. Conversely, the interaction was \emph{dis}assortative if $\omega_{rs} > \omega_{rr}$ and $\omega_{rs} > \omega_{ss}$. The core-periphery motif corresponded to the case where $\omega_{rr} > \omega_{rs} > \omega_{ss}$. In this case, communities $r$ and $s$ were labeled as the core and the periphery, respectively (note that $\omega_{ss} > \omega_{rs} > \omega_{rr}$ would also be considered a core-periphery motif, but with the roles reversed).

After classifying a community motif, we then passed the classification labels back to the nodes (neurons) that made up community $r$ and community $s$. That is, if the interaction between $r$ and $s$ was labeled ``assortative'', then all the neurons in $r$ and all the neurons in $s$ were assigned an assortative label. We repeated the assignment for all pairs of communities and calculated how frequently neurons were assigned one of the four classes (assortative, disassortative, core, and periphery).

\paragraph*{Determining the Rich Club}

To determine whether the signaling network contains a rich club we calculate the rich club coefficient $\Phi(k)$ across for every degree $k$. The rich club coefficient measures the density of connections among nodes with degree greater than $k$. We then compare $\Phi(k)$ for our network to the average $\Phi(k)_{rand}$ calculated across 1000 random networks in which the edges were randomly rewired while preserving the original degree of each node (using graph\_tool.generation.random\_rewire with the model `configuration'). While $\Phi(k)$ was determined to be statistically significantly larger than $\langle \Phi(k)_{rand} \rangle$ from $k=3$ onwards we concentrated on the two local maxima in  $\Phi(k)_{norm}$ at $k=25$ and $k=40$.

\paragraph*{Network Thresholding Controls}

We changed the thresholds of the networks to perform two different controls. 

We performed a comparison of the two networks, where both networks were thresholded to be more conservative in terms of their inclusion criteria for connections. For the signaling network, we used a threshold of $q = 0.01$ instead of $q = 0.05$, leading to a density of $d_s = 0.021$ (instead of $0.033$). To include a pair of neurons in the anatomical network, we required that they must have at least three synaptic contacts between them in at least one of the connectomes. This led to a density of $d_a = 0.028$ (instead of $0.092$). 

Additionally, in Extended Data Figures 12 and 13, we compared networks while enforcing their densities to be the same. We did so by thresholding the networks in either of two ways. For the anatomical network in Extended Data Figure 12, we required that for a pair of neurons to be anatomically connected the average synapses across the four connectome datasets be at least 1.25 synapses (instead of greater than zero), since this corresponds to a network connection density of $d_a = 0.033$ (comparable to the connection density of the signaling network $d_a = 0.033$). We then compared this thresholded anatomical network to the original signaling network.  We separately performed the converse adjustments in Extended Data Figure 13.  There, we required that the signaling network have a q-value threshold of $q <= 0.18$  (instead of $q<=0.05$) since this corresponds to a network connection density of $d_s = 0.091$ (comparable to the connection density of the anatomical network $d_s = 0.092$). We then compared this thresholded signaling network to the original anatomical network. 

\newpage

\renewcommand{\figurename}{Extended Data Figure}
\renewcommand{\tablename}{Extended Data Table}
\renewcommand{\thefigure}{\arabic{figure}}
\renewcommand{\thetable}{\arabic{table}}
\setcounter{figure}{0}
\setcounter{table}{0}

\newpage

\section*{Extended Data}

\begin{figure}[htbp]
\centering
\includegraphics[page=1, width=.63\linewidth]{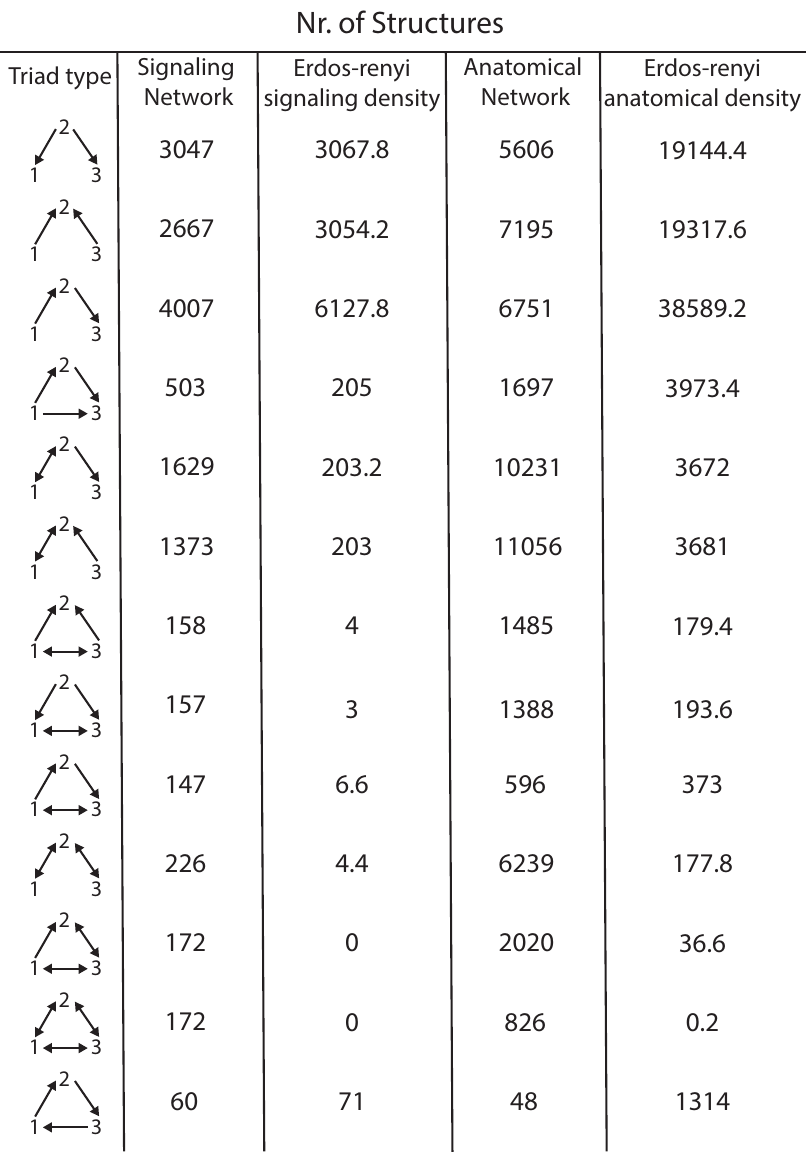}
\caption{ \textbf {The distribution of triad structures.} Occurrences of the 13 possible triad structures in the signaling and anatomical networks compared to the average from five random Erdos-renyi graphs with the same number of nodes and density as the signaling network (third column) and the anatomical network (fifth column)}  \label{fig:triads}
\end{figure}

\begin{table}[]
\centering
\begin{tabular}{|l|l|l|l|l|l|l|}
\hline
Network &
  \begin{tabular}[c]{@{}l@{}}Original \\ Signaling\end{tabular} &
  \begin{tabular}[c]{@{}l@{}}Original \\ Anatomy\end{tabular} &
  \begin{tabular}[c]{@{}l@{}}Conservative\\ Signaling\end{tabular} &
  \begin{tabular}[c]{@{}l@{}}Conservative \\ Anatomy\end{tabular} &
  \begin{tabular}[c]{@{}l@{}}Enforced density\\ Signaling\end{tabular} &
  \begin{tabular}[c]{@{}l@{}}Enforced density \\ Anatomy\end{tabular} \\ \hline
\begin{tabular}[c]{@{}l@{}}L/R neurons \\ same-community\\ odds\end{tabular} &
  2 &
  38 &
  1.3 &
  8 &
  2.43 &
  78 \\ \hline
\end{tabular}
\caption{How much more likely L/R pairs of neurons are to be in the same community than not depending on network thresholds.}
\label{tab:LR_table}
\end{table}

\begin{figure}[htbp]
\centering
\includegraphics[page=1, width=.85\linewidth]{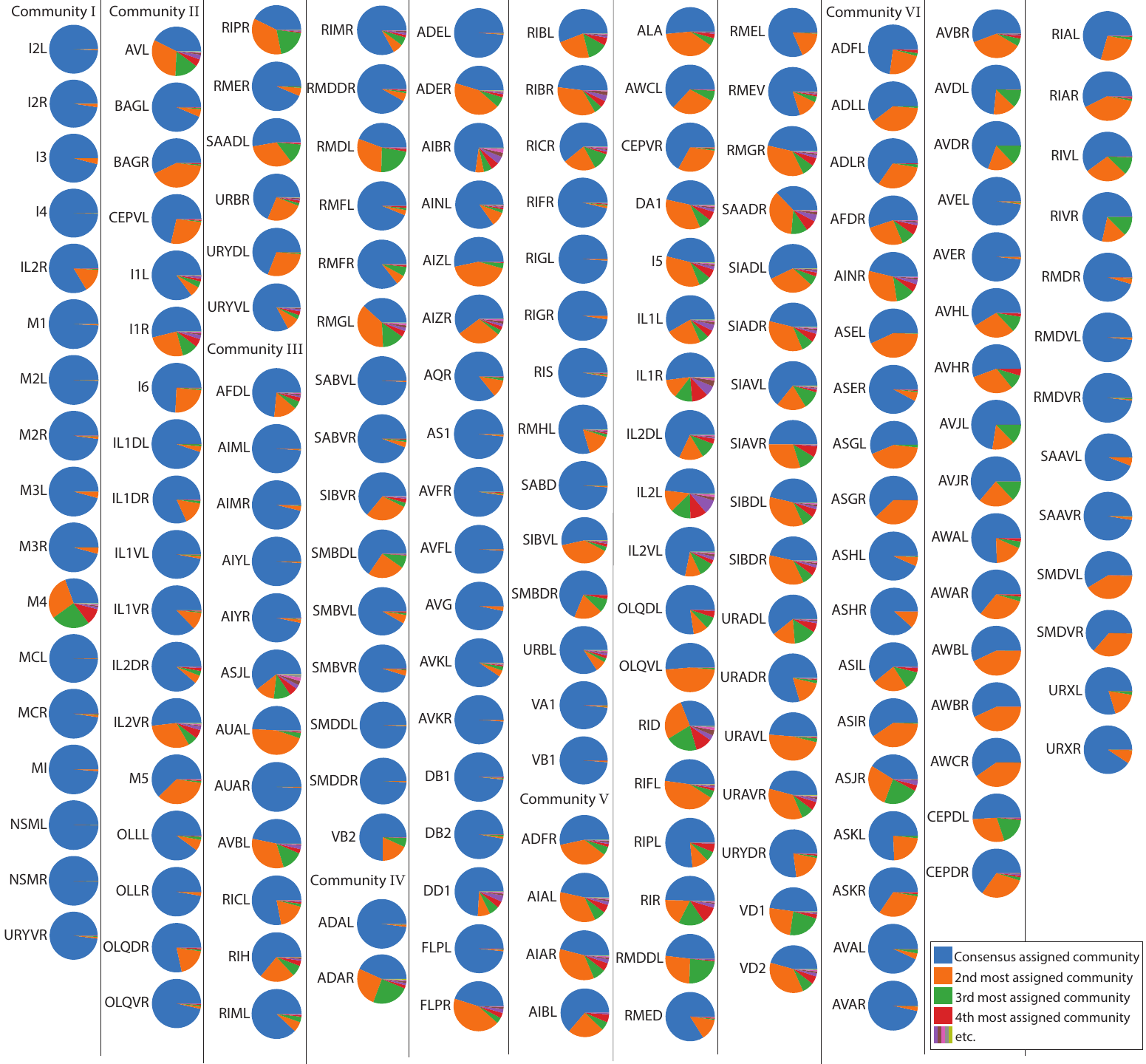}
\caption{ \textbf {Variability in community assignment by neuron.} Pie charts indicate how consistently neurons were assigned to the same community in the signaling network, organized by community. Blue corresponds to the fraction assigned to the consensus community assignment (largest fraction), orange the next most common community assignment then green etc. }  \label{fig:pies}
\end{figure}

\begin{figure}[htbp]
\centering
\includegraphics[page=1, width=.85\linewidth]{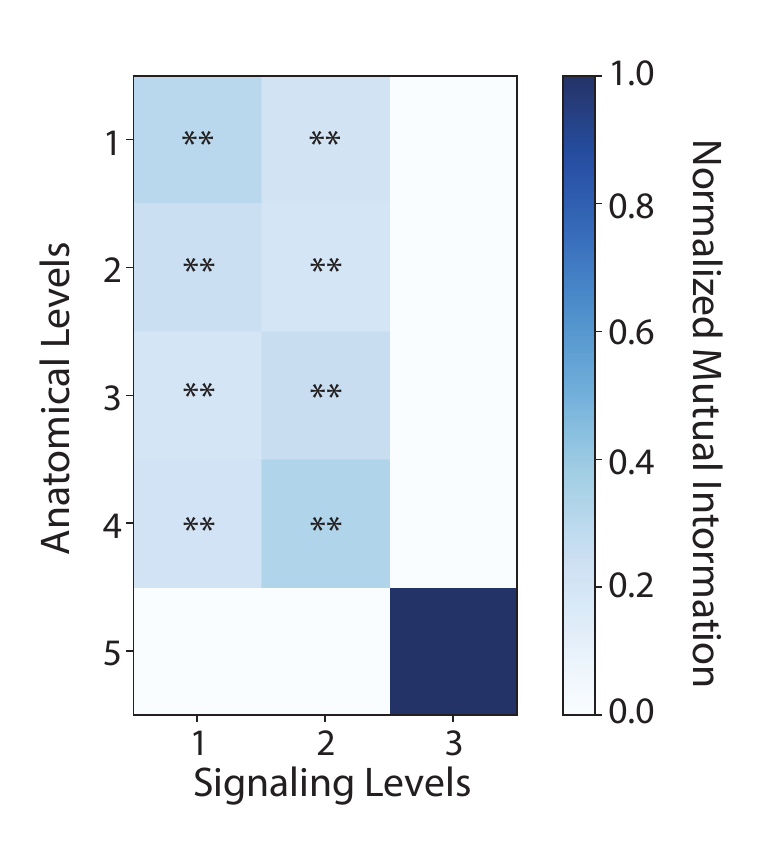}
\caption{ \textbf {Comparison of signaling and anatomical communities.} Normalized mutual information between the community assignment of neurons pairwise between hierarchical levels in the anatomical and signaling communities. ** indicates a FDR adjusted $p>0.05$ compared to randomly shuffled community assignments. The last community of both the anatomical and signaling networks only has one community that includes all of the neurons.}  \label{fig:comp_levels}
\end{figure}

\begin{figure}[htbp]
\centering
\includegraphics[page=1, width=.70\linewidth]{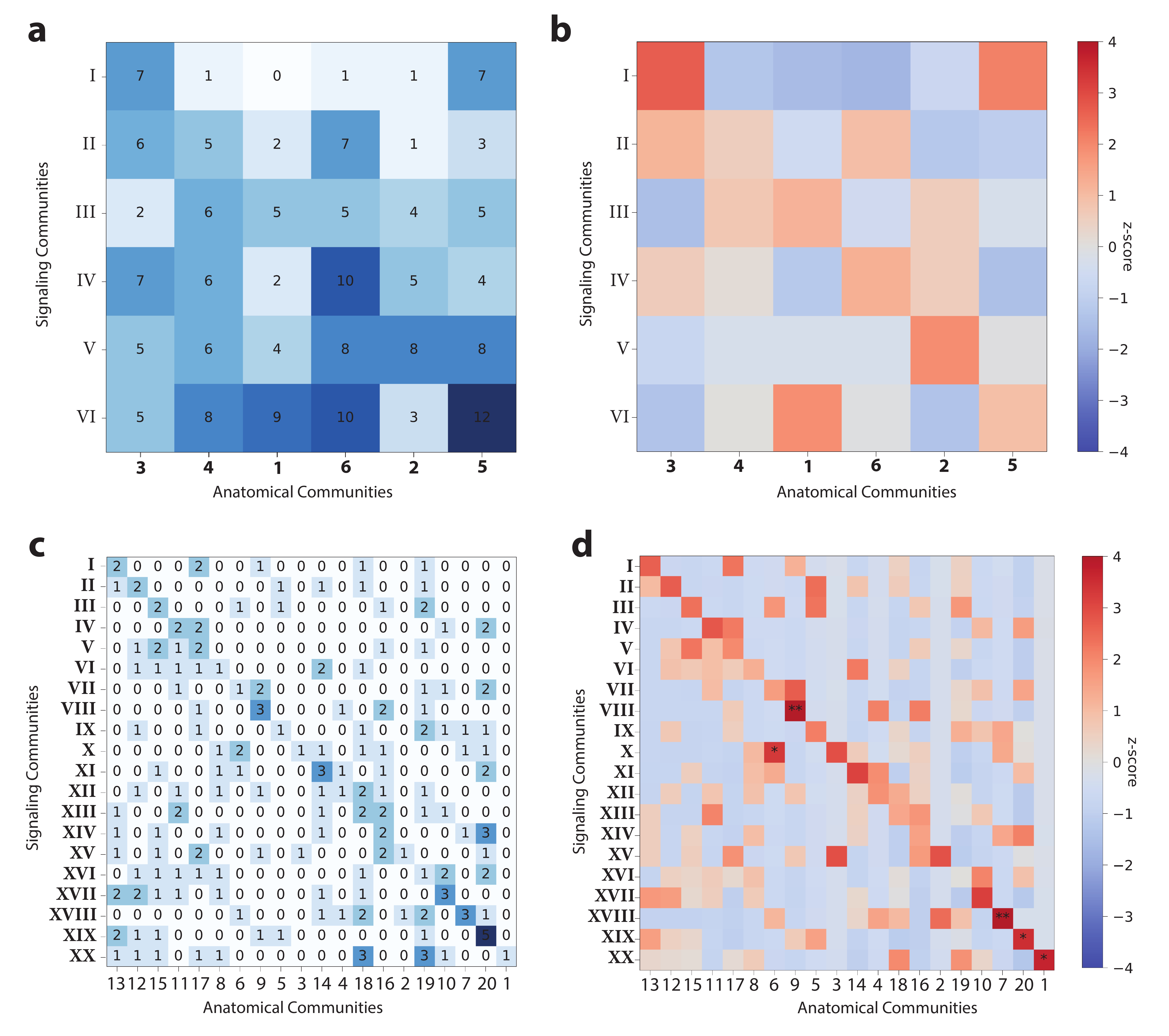}
\caption{ \textbf {Overlap between anatomical and signaling networks was not improved by forcing the number of communities to be the same. a)} Overlap between the anatomical and signaling communities. When partitioning the networks via stochastic block modeling, the anatomical network was forced to be partitioned into the same number of communities as the optimal partition for the signaling network (6 communities). \textbf{b)} Relative enrichment of
overlap between anatomy and signaling reported as a z-score for each anatomy-signaling community pair (* $p < 0.05$, ** $p < 0.01$,
p-values from the same distribution as z-score, FDR-adjusted). \textbf{c,d)} Same as a and b however, with the partitioning for the signaling network forced to be the same number of communities as the optimal partition for the anatomical network (20 communities). Anatomical communities in all sub-panels are organized by best-matching via a similarity matrix and the Hungarian method.}  \label{fig:same_blocks}
\end{figure}

\begin{figure}[htbp]
\centering
\includegraphics[page=1, width=.85\linewidth]{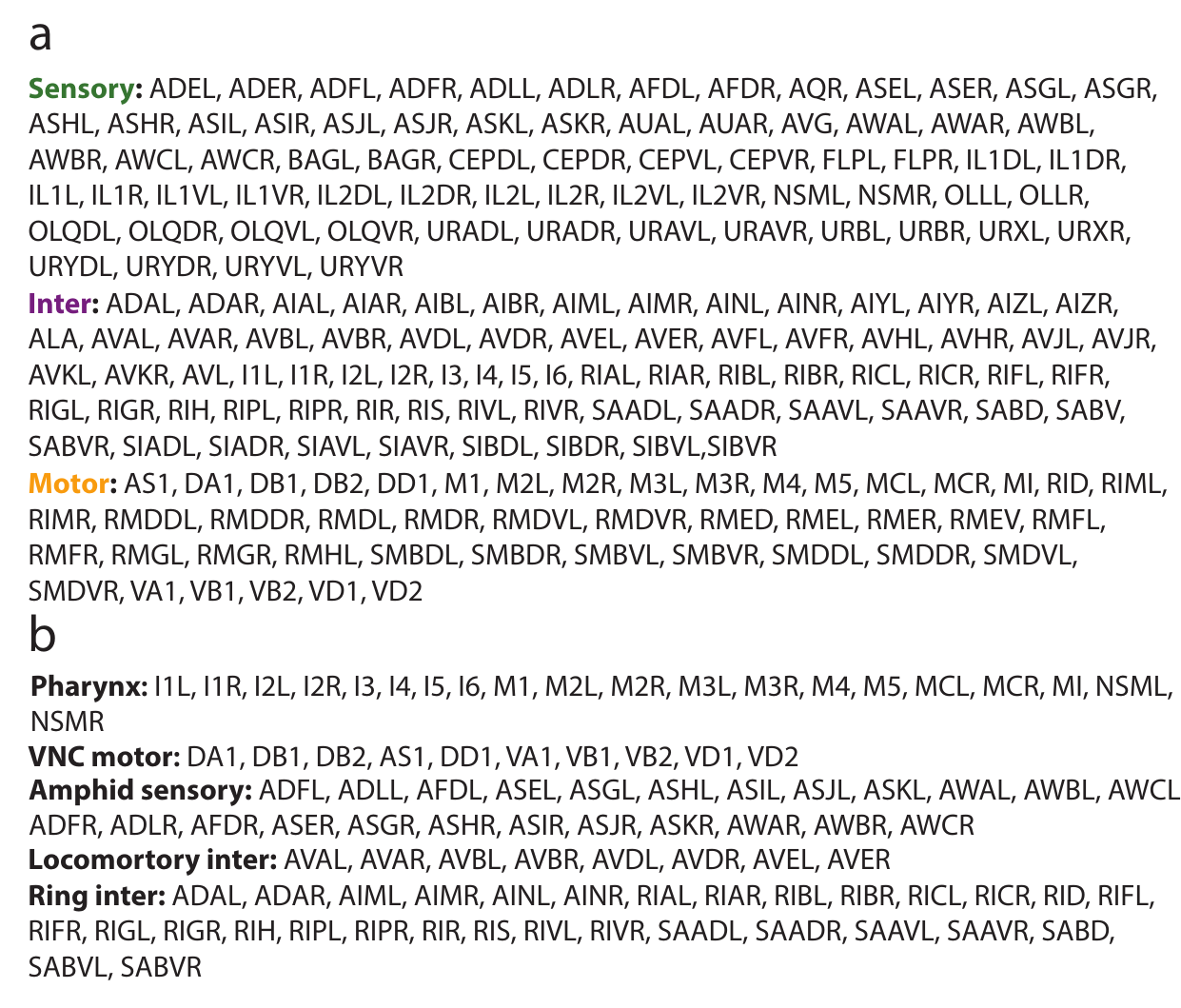}
\caption{ \textbf {Neuron type and role annotations. a)} Neurons in the different neuron types (sensory, inter- and motor neurons)   \textbf{b)} Neurons with different roles }  \label{fig:enrich_sup_nlist}
\end{figure}

\begin{figure}[t]
\centering
\includegraphics[page=1, width=\linewidth]{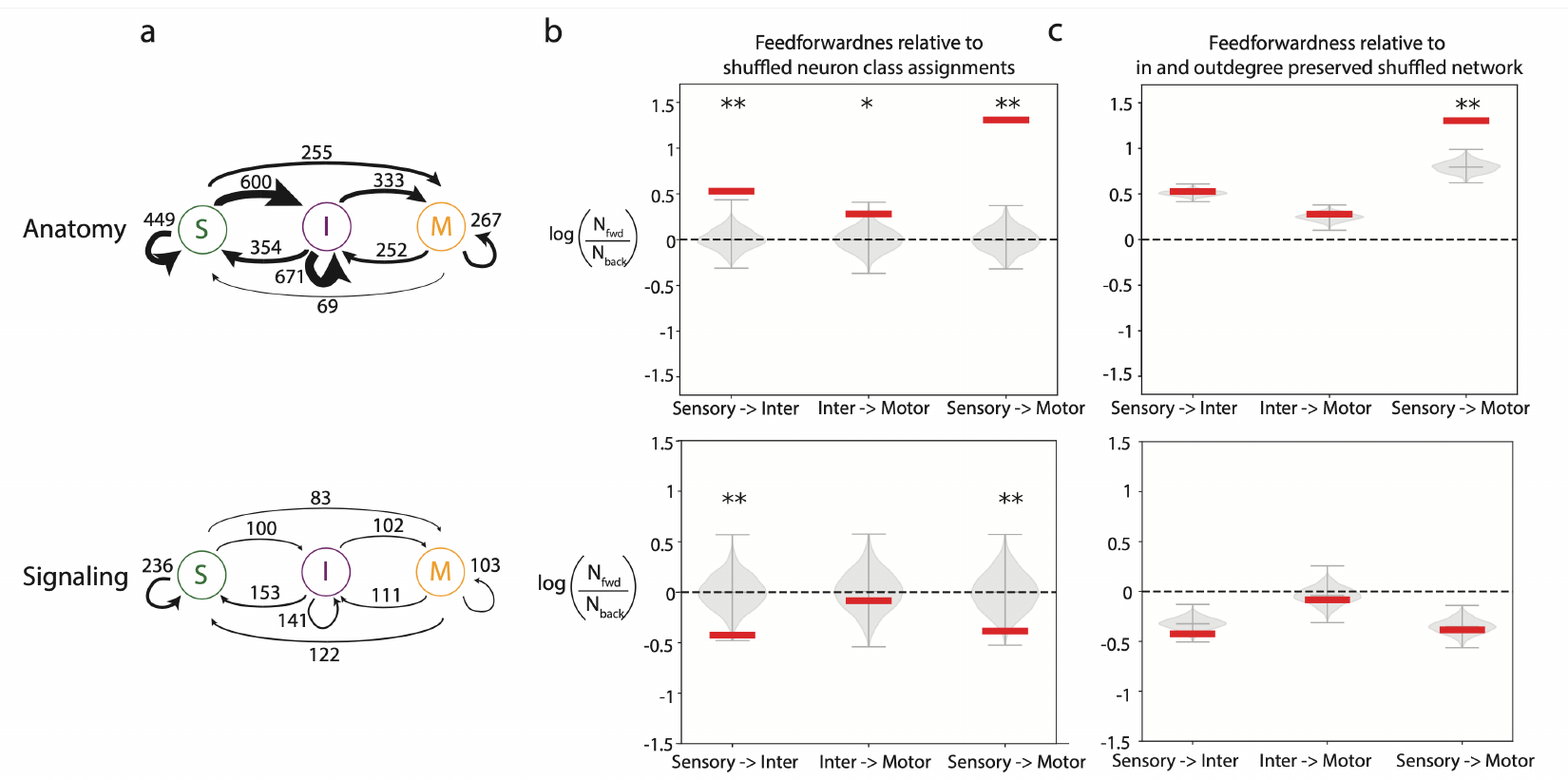}
\caption{ \textbf {Anatomy is net feed-forward along the sensory to inter- to motor neuron pathway while signaling is ambiguous. a)} Number of edges between sensory, inter-, and motor neurons for anatomy (top) and signaling (top). Arrow thicknessis  proportional to the number of edges along the arrow. \textbf{b)} log of the ratio of the number of edges forwards to backwards along the sensory $\rightarrow$ inter $\rightarrow $ motor pathway for each possible leg of the pathway: sensory $\rightarrow$ interneurons, inter- $\rightarrow$ motor neurons and sensory $\rightarrow$ motor neurons. Anything above $0$ indicates more forwards than backward edges implying feed-forwardness, anything below indicates more backward than forward edges implying feed-backwardness. Red indicates the true log of the ratio of the number of edges. Grey violin plots are derived from a control in which assignments of the labels "sensory", "inter" and "motor" are shuffled. \textbf{c)} Same as b) except the grey violin plots are derived instead from a control where the network is shuffled in a way that preserves both the total in and out-degree of each neuron. (* p<0.05, ** p<0.01, p-values calculated via a two-sided t-test)}  \label{fig:SIM2}
\end{figure}

 \begin{figure}[htbp]
\centering
\includegraphics[page=1, width=\linewidth]{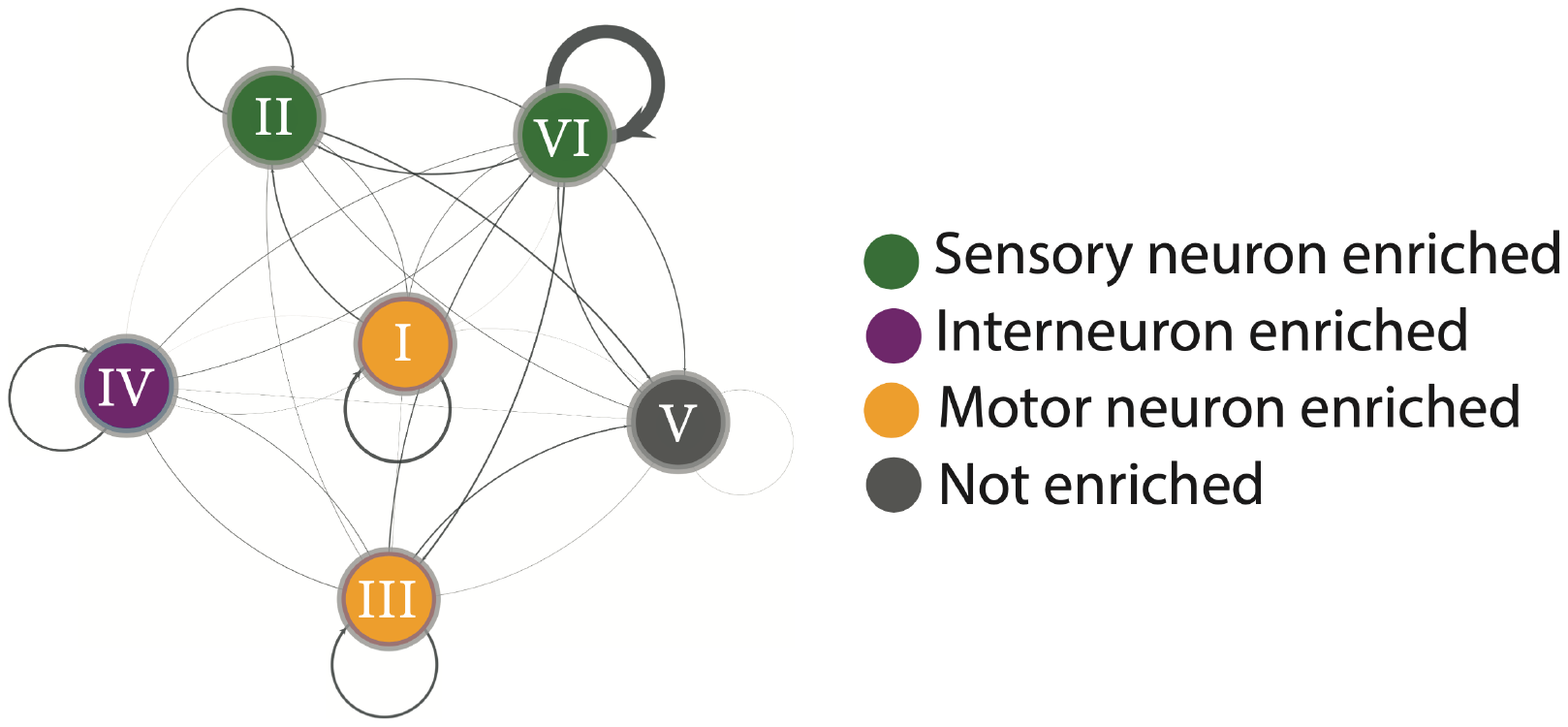}
\caption{\textbf{Network diagram of edges between the six communities.} Communities are colored by whether they are enriched for sensory (green), inter- (purple), motor neurons (yellow), or not enriched for any neuron type (grey). Arrow thickness indicates the fraction of all possible edges between nodes in the source community and the target community}  \label{fig:SIM}
\end{figure}

\begin{figure}[htbp]
\centering
\includegraphics[page=1, width=.85\linewidth]{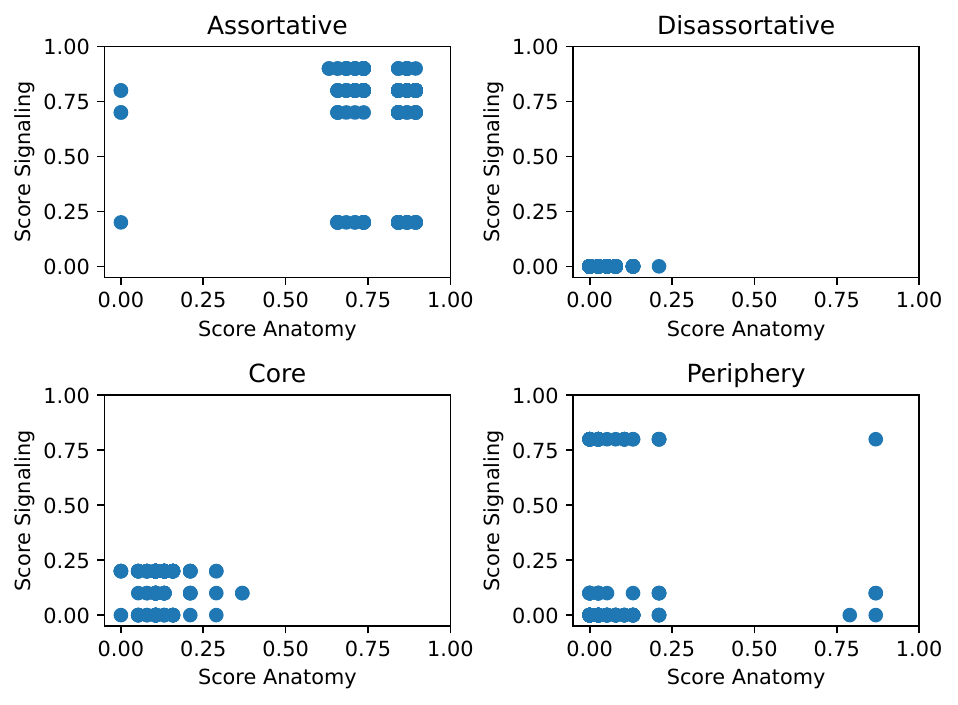}
\caption{ \textbf { Comparison of community motif score in anatomy and signaling.} Score assigned to each neuron for participating in assortative, disassortative, core, and periphery interactions in the anatomical network vs the signaling network}  \label{fig:enrich_sup}
\end{figure}

\begin{figure}[htbp]
\centering
\includegraphics[page=1, width=.60\linewidth]{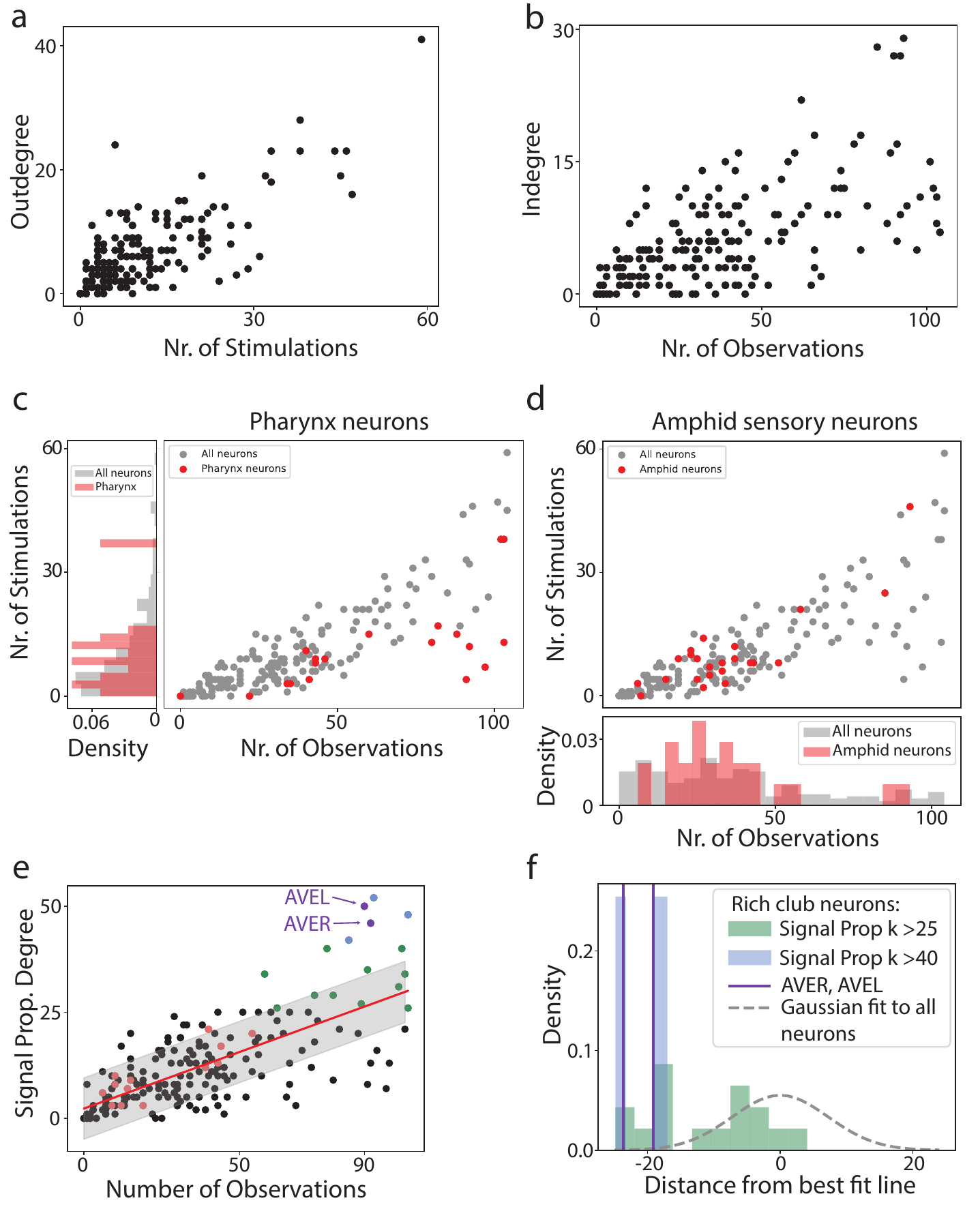}
\caption{ \textbf {Broadcaster, integrator and rich club results are not a trivial consequence of the number of stimulations and observations of neurons. a)} Scatter plot of out-degree and number of stimulations of all neurons. \textbf{b)} Scatter plot of the in-degree and the number of observations of all neurons \textbf{c)} Right panel: Histogram of the density of the number of stimulations of all neurons (grey) and pharynx neurons (red). Left panel: Scatter plot of the number of observations and the number of stimulations of all neurons (grey) and pharynx neurons (red). Pharynx neurons do not have a larger than expected number of stimulations compared to all other neurons (Kolmogorov-Smirnoff test p-value = 0.9)  \textbf{d)} Upper panel: Scatter plot of the number of observations and the number of stimulations of all neurons (grey) and amphid sensory neurons (red). Lower panel: Histogram of the density of the number of observations of all neurons (grey) and amphid neurons (red).  Amphid neurons do not have a larger than expected number of observations compared to all other neurons (Kolmogorov-Smirnoff test p-value = 0.38) \textbf{e)} Correlation between the number of observations and signal propagation degree (pearsons correlation $r = 0.71$). Green: signaling rich club neurons of $k>25$, blue: signaling rich club of neurons $k>40$, red: anatomical rich club neuron, purple: neurons in both signaling and anatomical rich clubs. \textbf{f)} Histogram of neurons distance from the best-fit line in e). Dashed grey line shows the Gaussian fit to the histogram of all neurons. Blue and green show the histograms of the signaling rich club neurons with $k>40$ and $k>25$ respectively. Purple bars show neurons AVEL/R.  }  \label{fig:intbroad}
\end{figure}

\begin{figure}[htbp]
\centering
\includegraphics[page=1, width=.60\linewidth]{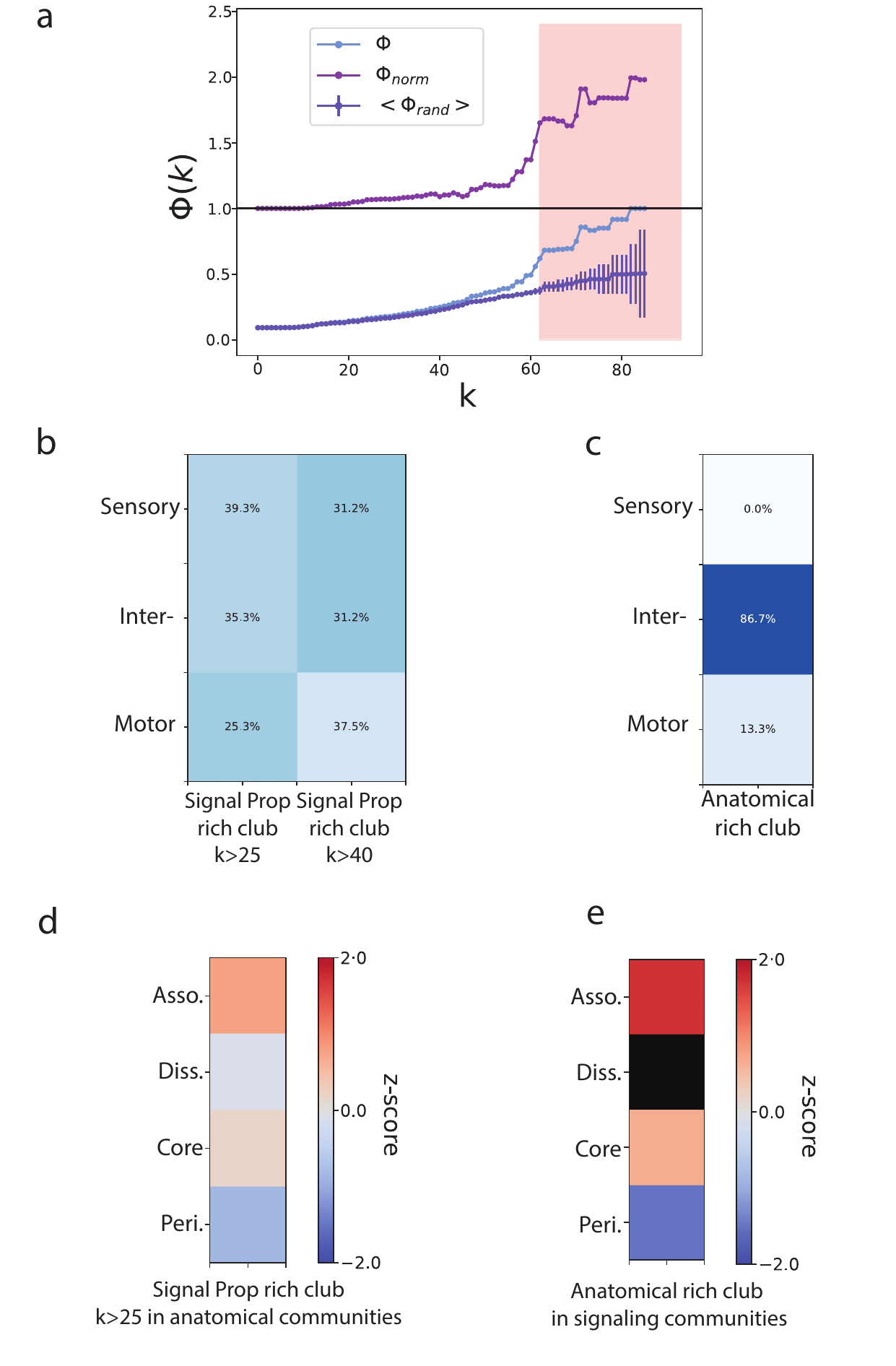}
\caption{ \textbf {Rich Club. a)} The anatomical network has a rich club. Light blue: rich club coefficient of the anatomical connectome, dark blue: average rich club coefficient of 100 randomized anatomical networks (edges rewired randomly but the degree sequence remains the same) error bars represent standard deviation, purple: normalized rich club coefficient. \textbf{b)} Sensory, inter-, and motor neuron makeup of the signal propagation rich club. \textbf{c)} Sensory, inter-, and motor neuron makeup of the anatomical rich club. \textbf{d)} Enrichment
(z-score) of neurons belonging to the signaling rich club for assortative, disassortative, and core/periphery interactions in the anatomical network (* p<0.05, ** p<0.01, p-values
FDR-adjusted).  \textbf{e)} Enrichment
(z-score) of neurons belonging to the anatomical rich club for assortative, disassortative, and core/periphery interactions in the signaling network (* p<0.05, ** p<0.01, p-values
FDR-adjusted)  }  \label{fig:Rich_sup}
\end{figure}

\begin{figure}[htbp]
\centering
\includegraphics[page=1, width=.70\linewidth]{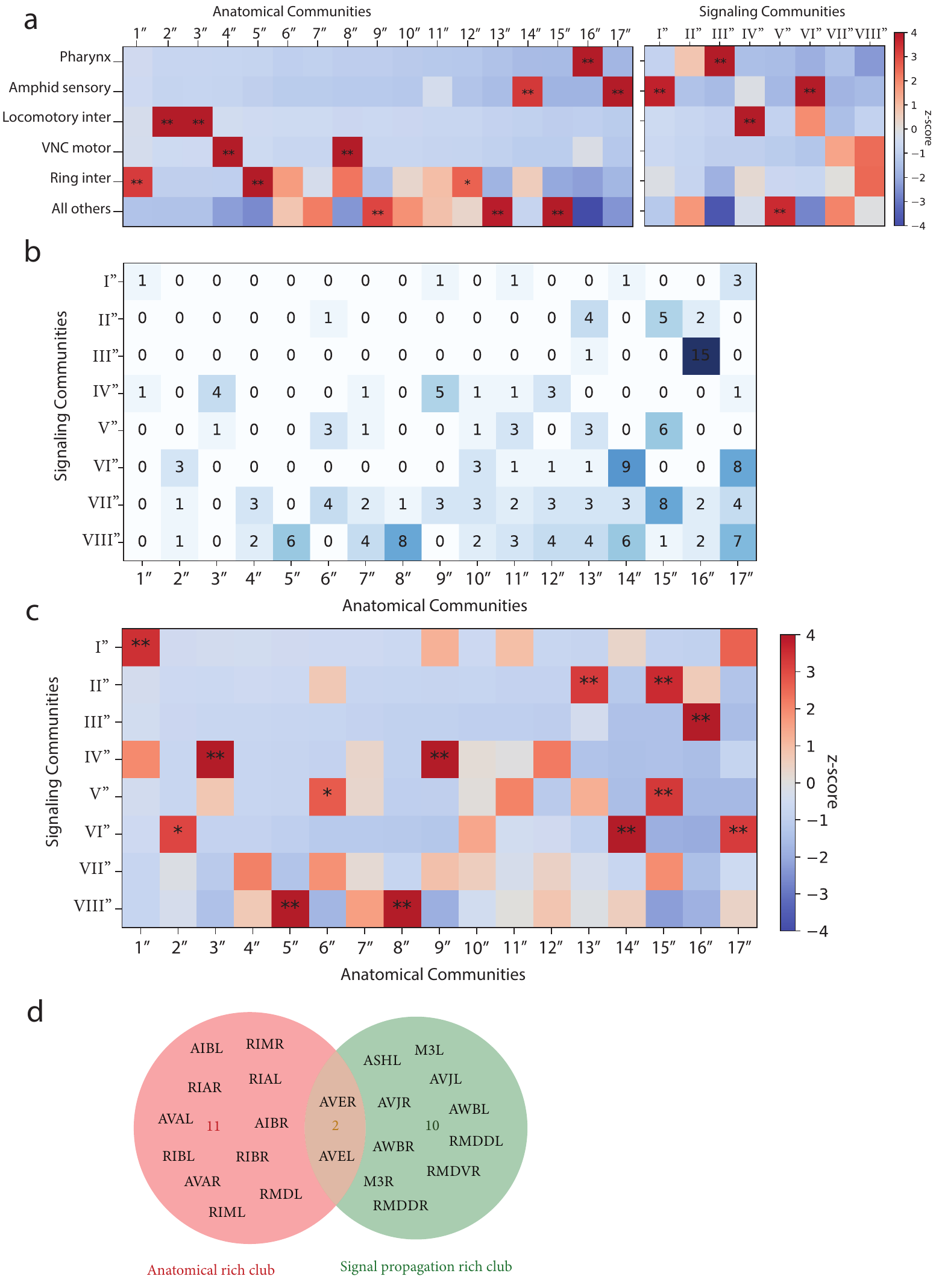}
\caption{ \textbf {A more conservative thresholding of both networks yields similar results.} \textbf{a)} Enrichment of thresholded anatomical and signaling communities for neuron role (* $p<0.05$, ** $p<0.01$, p-values FDR-adjusted). \textbf{b)} The anatomical rich club is still non-overlapping with the signaling rich club, other than neurons AVEL/R.  \textbf{c)}  Overlap (counts)
between membership of thresholded anatomical and signal propagation communities. \textbf{d)} Relative enrichment of overlap is reported as a z-score for each thresholded anatomy-signaling community pair  (* $p<0.05$, ** $p<0.01$, p-values from same distribution as z-score, FDR-adjusted). }  \label{fig:conservative_thresh}
\end{figure}

\begin{figure}[htbp]
\centering
\includegraphics[page=1, width=.70\linewidth]{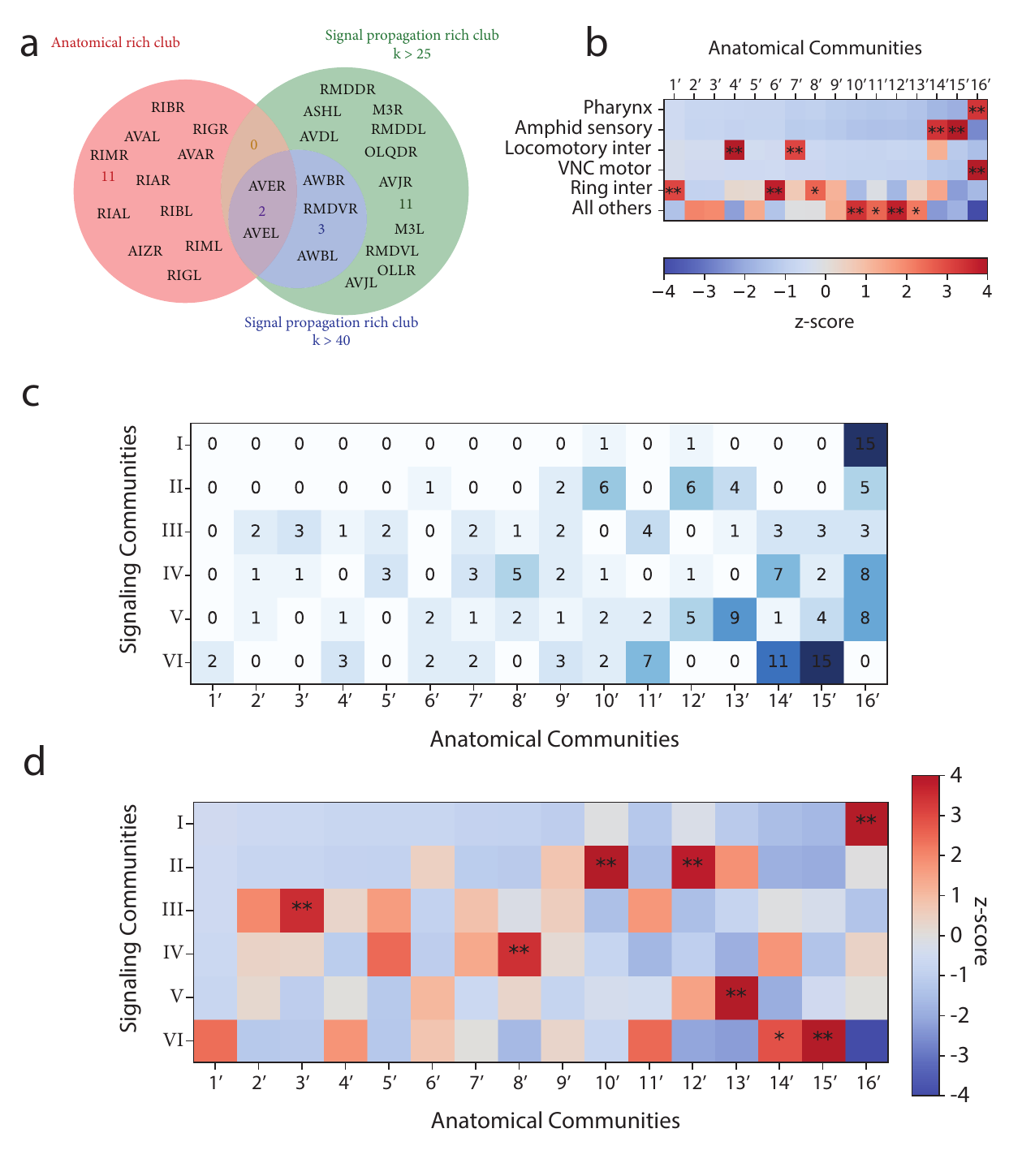}
\caption{ \textbf {Adjusting the threshold of the anatomical connectome to match the density of the signaling network yields similar results.} The anatomical network was thresholded based on the number of synapses needed for there to be an edge between two neurons to closely match the density of the signaling network $d_a = 0.033$. The thresholded anatomical network was optimally partitioned into 16 communities. \textbf{a)} The anatomical rich club is still non-overlapping with the signaling rich club, other than neurons AVEL/R. The anatomical rich club consists mostly of the same neurons as originally (9 out of the 15 original neurons are the same, with 4 new neurons). \textbf{b)} Enrichment of anatomical communities for neuron role (* $p<0.05$, ** $p<0.01$, p-values FDR-adjusted). \textbf{c)}  Overlap (counts)
between membership of anatomical and signal propagation communities. \textbf{d)} Relative enrichment of overlap is reported as a z-score for each anatomy-signaling community pair  (* $p<0.05$, ** $p<0.01$, p-values from same distribution as z-score, FDR-adjusted). }  \label{fig:aconn_thresh}
\end{figure}

\begin{figure}[htbp]
\centering
\includegraphics[page=1, width=.70\linewidth]{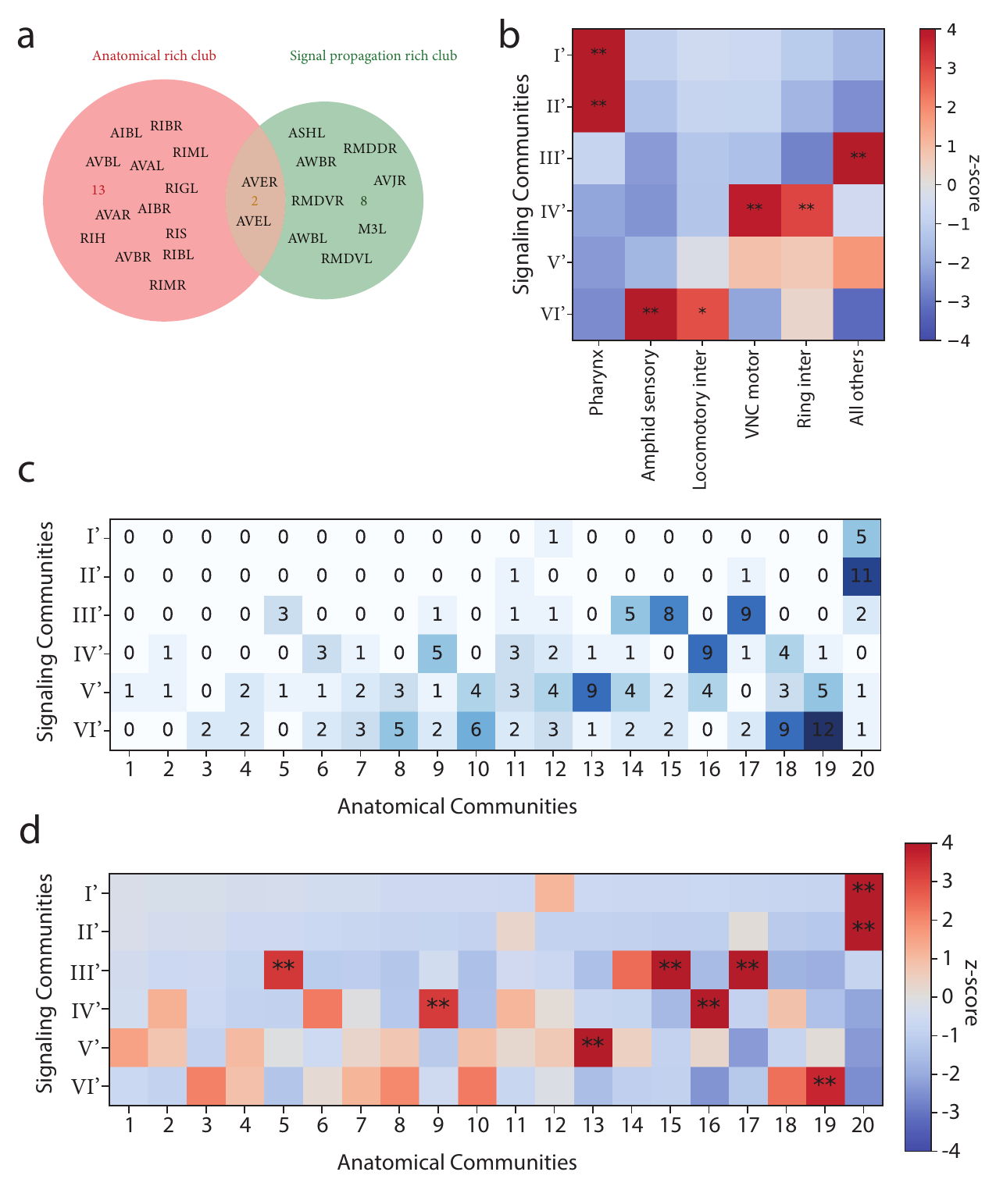}
\caption{ \textbf {Adjusting the threshold of the signaling network to match the density of the anatomical network yields similar results.} The signaling network was thresholded based on the maximum q-value of a connection between two neurons needed for there to be an edge between two neurons to closely match the density of the anatomical network $d_s = 0.091$. The thresholded signaling network was also optimally partitioned into 6 communities. \textbf{a)} The signaling rich club is still non-overlapping with the anatomical rich club, other than neurons AVEL/R. All neurons in the signaling rich club were also a part of the original rich club. \textbf{b)} Enrichment of signaling communities for neuron role (* $p<0.05$, ** $p<0.01$, p-values FDR-adjusted). \textbf{c)}  Overlap (counts)
between membership of anatomical and signal propagation communities. \textbf{d)} Relative enrichment of overlap is reported as a z-score for each anatomy-signaling community pair  (* $p<0.05$, ** $p<0.01$, p-values from same distribution as z-score, FDR-adjusted). }  \label{fig:fconn_thresh}
\end{figure}

\begin{figure}[htbp]
\centering
\includegraphics[page=1, width=.70\linewidth]{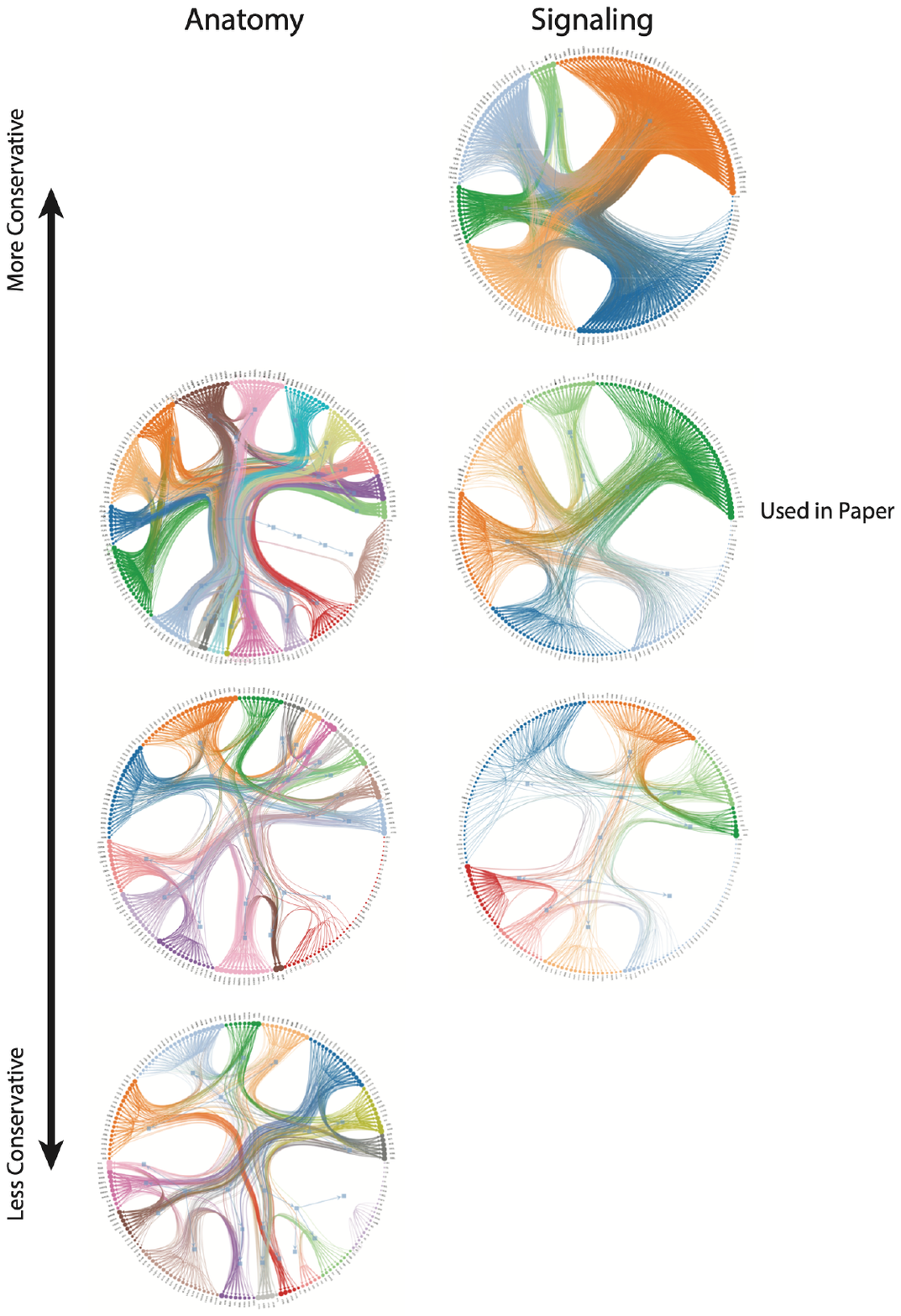}
\caption{ \textbf {Modular structure at different thresholds.} Circular dendograms showing community assignments at different hierarchical levels for anatomy (left) and signaling (right) at the different binarization thresholds used in the paper and in the controls.}  \label{fig:thresh_networks}
\end{figure}

\end{document}